\documentclass[11pt,a4paper]{article}

\usepackage{jheppub}

\usepackage{amsmath}

\usepackage{graphicx}

\usepackage{url}

\usepackage{epsfig}

\usepackage{feynmf}

\newcommand{\sq}{{\tilde{q}}}

\newcommand{\sqb}{{\bar{\tilde{q}}}}

\newcommand{\gl}{{\tilde{g}}}

\renewcommand{\d}{\mathrm{d}}

\newcommand{\as}{\alpha_{\rm s}}

\newcommand{\ashat}{\hat\alpha_{\rm s}}

\newcommand{\MSbar}{\rm{MS}\hspace{-17pt}\overline{\phantom{tmt}}\hspace{1pt}}

\title{NNLL resummation for squark and gluino production at the LHC}

\author[a]{Wim Beenakker,}

\author[b]{Christoph Borschensky,}

\author[c]{Michael Kr\"amer,}

\author[b]{Anna Kulesza,}

\author[d]{Eric Laenen,}

\author[b]{Vincent Theeuwes,}

\author[e]{Silja Thewes}

\affiliation[a]{Theoretical High Energy Physics, IMAPP, Faculty of Science, Mailbox 79, Radboud University Nijmegen, P.O. Box 9010, NL-6500 GL Nijmegen,\\
Institute of Physics, University of Amsterdam, Amsterdam, The Netherlands}

\affiliation[b]{Institute for Theoretical Physics, WWU M\"unster, D-48149 M\"unster, Germany}

\affiliation[c]{Institute for Theoretical Particle Physics and Cosmology, RWTH Aachen University D-52056 Aachen, Germany}

\affiliation[d]{ITFA, University of Amsterdam, Science Park 904, 1018 XE, Amsterdam,\\
ITF, Utrecht University, Leuvenlaan 4, 3584 CE Utrecht,\\
Nikhef Theory Group, Science Park 105, 1098 XG Amsterdam, The Netherlands}

\affiliation[e]{DESY, Theory Group, Notkestrasse 85, D-22603 Hamburg, Germany}

\emailAdd{W.Beenakker@science.ru.nl}

\emailAdd{borschensky@uni-muenster.de}

\emailAdd{mkraemer@physik.rwth-aachen.de}

\emailAdd{anna.kulesza@uni-muenster.de}

\emailAdd{t45@nikhef.nl}

\emailAdd{vthee\_01@uni-muenster.de}

\emailAdd{silja.christine.thewes@desy.de}

\abstract{We perform the resummation of soft-gluon emissions for squark and gluino production at next-to-next-to-leading logarithmic (NNLL) accuracy. We include also the one-loop hard matching coefficients as well as Coulomb corrections to second order, using Mellin-moment methods. We study the characteristics of this resummation in detail for a centre-of-mass (CM) energy of 8 TeV at the LHC, and for squark and gluino masses up to 2.5 TeV. We find significant enhancing effects for all four processes of squark- and gluino-pair production. Scale dependence is generally reduced compared to NLL resummation, except for gluino-pair production where we find a moderate enhancement.}


\begin{document}
\begin{flushright}
MS-TP-14-16\\TTK-14-08\\NIKHEF-2014-009\\ITP-UU-14/11\\
\end{flushright}
\maketitle

\begin{fmffile}{feyngraphs}

%feynmfstuff

\fmfset{dot_size}{1thick}

\fmfset{curly_len}{1.5mm}

\fmfset{arrow_len}{3mm}

\fmfpen{thin}

\setlength{\unitlength}{1mm}

\section{Introduction}

\label{s:intro}

Supersymmetry (SUSY)~\cite{Golfand:1971iw,Wess:1974tw} is one of the most promising extensions of the Standard Model, which can offer a solution to the hierarchy problem, result in gauge unification and provide a dark-matter candidate. In order for SUSY to be able to accommodate these solutions its scale needs to be comparable to the weak scale, leading to supersymmetric particles with masses near the TeV range. These particles, especially the coloured ones (squarks ($\sq$) and gluinos ($\gl$)), could be within range of the Large Hadron Collider (LHC). The current lower limit for the masses of the coloured supersymmetric particles has been set to around or above 1~TeV, depending on the model~\cite{Aad:2012rz,Chatrchyan:2012jx,Aad:2012ona,Chatrchyan:2012gq,Chatrchyan:2013lya}.

In the context of the Minimal Supersymmetric Standard Model (MSSM)~\cite{Nilles:1983ge,Haber:1984rc} with R-parity conservation, supersymmetric particles are formed in pairs in collisions of two hadrons $h_1$ and $h_2$. Squarks and gluinos can then be produced in the following manner:

\[h_1h_2 \;\to\; \sq\sq,\,\sq\sqb,\,\sq\gl,\,\gl\gl+X.\]
Here and in the rest of this paper the chiralities of the squarks $\sq=(\sq_L,\sq_R)$ will be suppressed, nor will we explicitly state the charge conjugated process.

Accurate theoretical predictions for inclusive production of squarks and gluinos are needed in order to improve exclusion limits and, should supersymmetry be found, more accurately study the masses and properties of the particles~\cite{Baer:2007ya,Dreiner:2010gv}. These predictions can be improved by including higher-order QCD corrections. The next-to-leading order (NLO) corrections have already been known for some time~\cite{Beenakker:1994an,Beenakker:1995fp,Beenakker:1996ch, Beenakker:1997ut}. A significant contribution to the NLO corrections comes from the region near threshold, where the partonic center-of-mass energy is close to the kinematic restriction for the on-shell production of these particles, i.e. $s \geq 4 m_{av}^2$ with $m_{av}$ being the average mass of the two produced particles. In this region the dominant corrections originate from the soft-gluon emission off the initial and final state as well as the exchange of a gluon between the slowly moving heavy final-state particles, giving rise to the Coulomb corrections. The soft-gluon corrections can be taken into account to all orders in perturbation theory using threshold resummation techniques~\cite{Sterman:1986aj,Catani:1989ne,Bonciani:1998vc,Contopanagos:1996nh,Kidonakis:1998bk,Kidonakis:1998nf}.

Threshold resummation has been performed for all MSSM squark and gluino production processes at next-to-leading logarithm (NLL) accuracy~\cite{Kulesza:2008jb,Kulesza:2009kq,Beenakker:2009ha,Beenakker:2010nq,Beenakker:2011fu,Beenakker:2013mva}. For both squark-antisquark and gluino-pair production, in addition to the soft-gluon resummation, the Coulomb corrections have been resummed both by using a Sommerfeld factor~\cite{Kulesza:2009kq} and by using the framework of effective field theories~\cite{Beneke:2010da,Falgari:2012hx}. Furthermore, the dominant next-to-next-to-leading order (NNLO) corrections, coming from the resummed next-to-next-to-leading logarithm (NNLL) expression, have been calculated for squark-antisquark and gluino-pair production~\cite{Langenfeld:2009eg,langenfeldgluino}. For squark-antisquark production, soft-gluon emissions have been resummed to NNLL level~\cite{Beenakker:2011sf}, and the same has been achieved for gluino-pair production~\cite{Pfoh:2013iia}. Recently, NNLL predictions have been obtained for stop-pair production in the framework of soft-collinear effective theory~\cite{Broggio:2013cia}. The finite-width effects have also been studied for squark and gluino production processes in~\cite{Falgari:2012sq}.

In this paper, to illustrate the effects of NNLL resummation, we consider all four pair-production processes of squarks and gluinos at the LHC with $\sqrt S=8$ TeV. We examine squark and gluino masses up to 2.5~TeV, and use the Mellin-moment-space approach. These settings allow us to examine the resummation technique for processes that carry large colour factors in an extreme setting, i.e.~very close to threshold. In the near future we plan to upgrade the NLL resummation code {\sc {NLL-fast}}~\cite{nllfast} to NNLL level, which will also involve producing NNLL resummed results for the LHC at $\sqrt S=13$ TeV.

The paper is structured as follows. In the next section we briefly review the theoretical expressions for NNLL-resummed cross sections applied to the particular case of pair production of squarks and gluinos. In Section 3 we discuss the numerical predictions for the NNLL-resummed cross sections matched to the approximate NNLO results. Section 4 contains our conclusions. 
The appendices contain the one- and two-loop Coulomb corrections in Mellin-moment space.

%%%%%%%%%%%%%%%%%%%%%%%%%%%%%%%%%%%%%%%%%%%

\section{NNLL resummation}

%%%%%%%%%%%%%%%%%%%%%%%%%%%%%%%%%%%%%%%%%%%

\label{s:theory}

Before the calculation of the NNLL resummation, we will first briefly review the formalism of threshold resummation for the pair production of squarks and gluinos. The inclusive hadronic cross section for the production of particles $k$ and $l$, $\sigma_{h_1h_2\to kl}$, can be written in terms of the partonic cross section, $\sigma_{ij\to kl}$, in the following manner
\begin{multline}
  \label{eq:hadr-cross}
  \sigma_{h_1 h_2 \to kl}\bigl(\rho, \{m^2\}\bigr) 
  \;=\; \sum_{i,j} \int d x_1 d x_2\,d\hat{\rho}\;
        \delta\left(\hat{\rho} - \frac{\rho}{x_1 x_2}\right)\\
        \times\,f_{i/h_{1}}(x_1,\mu^2 )\,f_{j/h_{2}}(x_2,\mu^2 )\,
        \sigma_{ij \to kl}\bigl(\hat{\rho},\{ m^2\},\mu^2\bigr)\,,
\end{multline}
where $\{m^2\}$ denotes all the masses entering the calculation, $i$ and $j$ are the initial-state parton flavours, $f_{i/h_{1}}$ and $f_{j/h_{2}}$ are the parton distribution functions, $\mu$ is the common factorisation and renormalisation scale, $x_1$ and $x_2$ are the momentum fractions of the partons inside the hadrons $h_1$ and $h_2$, and $\rho$ and $\hat{\rho}$ are the hadronic and partonic threshold variables respectively. The threshold for the production of two final-state particles $k$ and $l$ with masses $m_k$ and $m_l$ corresponds to a hadronic center-of-mass energy squared of $S={(m_k+m_l)}^2$. Therefore we define the hadronic threshold variable $\rho$, measuring the distance from threshold in terms of a quadratic energy fraction, as

\[\rho \;=\; \frac{{(m_k+m_l)}^2}{S}\,.\]

In the threshold region, the dominant contributions to the higher-order QCD corrections due to soft-gluon emission have the general form
\begin{equation}
\alpha_{\rm s}^n \log^m\!\beta^2\ \ , \ \ m\leq 2n 
\qquad {\rm \ with\ } \qquad 
\beta^2 \,\equiv\, 1-\hat{\rho} \,=\, 1 \,-\, \frac{4(m_{av})^2}{s}\,,
\label{eq:beta}
\end{equation}
where $s=x_1x_2S$ is the partonic center-of-mass energy squared, $\alpha_{\rm s}$ is the strong coupling and $m_{av}=(m_k+m_l)/2$ is the average mass of the final-state particles $k$ and $l$. We perform the resummation of the soft-gluon emission after taking the Mellin transform (indicated by a tilde) of the cross section:
\begin{align}
  \label{eq:Mellin-transf}
  \tilde\sigma_{h_1 h_2 \to kl}\bigl(N, \{m^2\}\bigr) 
  &\equiv \int_0^1 d\rho\;\rho^{N-1}\;
           \sigma_{h_1 h_2\to kl}\bigl(\rho,\{ m^2\}\bigr) \nonumber\\
  &=      \;\sum_{i,j} \,\tilde f_{i/{h_1}} (N+1,\mu^2)\,
           \tilde f_{j/{h_2}} (N+1, \mu^2) \,
           \tilde{\sigma}_{ij \to kl}\bigl(N,\{m^2\},\mu^2\bigr)\,.
\end{align}
The logarithmically enhanced terms now take the form of $\alpha_{\rm s}^n \log^m\!N$, $m\leq 2n$, where the threshold limit $\beta \to 0$ corresponds to $N \to \infty$. The all-order summation of such logarithmic terms follows from the near-threshold factorisation of the cross section into functions that each capture the contributions of classes of radiation effects: hard, collinear and wide-angle soft radiation \cite{Sterman:1986aj,Catani:1989ne,Bonciani:1998vc,Contopanagos:1996nh,Kidonakis:1998bk,Kidonakis:1998nf}. Near threshold the resummed partonic cross section takes the form:
\begin{align}
  \label{eq:resummed-cross}
  \tilde{\sigma}^{\rm (res)} _{ij\to kl}\bigl(N,\{m^2\},&\mu^2\bigr) 
  =\sum_{I}\,
      \tilde\sigma^{(0)}_{ij\to kl,I}\bigl(N,\{m^2\},\mu^2\bigr)\, 
C_{ij\to kl,I}(N,\{m^2\},\mu^2)\nonumber\\
  & \times\,\Delta_i (N+1,Q^2,\mu^2)\,\Delta_j (N+1,Q^2,\mu^2)\,
     \Delta^{\rm (s)}_{ij\to kl,I}\bigl(Q/(N\mu),\mu^2\bigr)\,,
\end{align}
where we have introduced the hard scale $Q^2=4m_{av}^2$. The soft radiation is coherently sensitive to the colour structure of the hard process from which it is emitted \cite{Bonciani:1998vc,Contopanagos:1996nh,Kidonakis:1998bk,Kidonakis:1998nf,Botts:1989kf,Kidonakis:1997gm}. At threshold, the resulting colour matrices become diagonal to all orders by performing the calculation in the $s$-channel colour basis \cite{Beneke:2009rj,Kulesza:2008jb,Kulesza:2009kq}. The different contributions then correspond to different irreducible representations $I$. Correspondingly, $\tilde\sigma^{(0)}_{ij\to kl,I}$ in equation~\eqref{eq:resummed-cross} are the colour decomposed leading-order (LO) cross sections. The collinear radiation effects are summed into the functions $\Delta_i$ and $\Delta_j$ and the wide-angle soft radiation is described by $\Delta^{\rm (s)}_{ij\to kl,I}$. The radiative factors can then be written as
\begin{equation}
  \label{eq:NNLL-expa}
\Delta_i\Delta_j\Delta^{\rm(s)}_{ij\to kl,I}
  \;=\; \exp\Big[L g_1(\alpha_{\rm s}L) + g_2(\alpha_{\rm s}L) + \alpha_{\rm s}g_3(\alpha_{\rm s}L) + \ldots \Big]  \,.
\end{equation}
This exponent contains all the dependence on large logarithms $L=\log N$. The leading logarithmic approximation (LL) is represented by the $g_1$ term alone, whereas the NLL approximation requires additionally including the $g_2$ term. Similarly, the $g_3$ term is needed for the NNLL approximation. The customary expressions for the $g_1$ and $g_2$ functions can be found in e.g.~\cite{Kulesza:2009kq} and the one for the NNLL $g_3$ function in e.g.~\cite{Beenakker:2011sf}.

The matching coefficients $C_{ij\to kl,I}$ in~\eqref{eq:resummed-cross} collect non-logarithmic terms as well as logarithmic terms of non-soft origin in the Mellin moments of the higher-order contributions. The coefficients  $C_{ij\to kl,I}$ factorise into a part that contains the Coulomb corrections and a part containing hard contributions~\cite{Beneke:2010da} 
\begin{equation}
C_{ij\to kl,I}= (1+ \frac{\alpha_{\rm s}}{\pi}{\cal C}_{ij\to kl,I}^{\rm Coul,(1)}+\frac{\alpha_{\rm s}^2}{\pi^2}{\cal C}_{ij\to kl,I}^{\rm Coul,(2)}+\dots)(1+ \frac{\alpha_{\rm s}}{\pi}{\cal C}_{ij\to kl,I}^{(1)}+ \frac{\alpha_{\rm s}^2}{\pi^2}{\cal C}_{ij\to kl,I}^{(2)}+\dots)\,.
\label{eq:factCcoeff}
\end{equation}
Apart from the terms of ${\cal O}(\alpha_s)$, which need to be included in $C_{ij\to kl,I}$ when performing resummation at NNLL, some of the ${\cal O}(\alpha_s^2)$ terms are also known and can be included in the numerical calculations. Expanding \eqref{eq:factCcoeff} we have 
\begin{align}
C^{\rm NNLL}_{ij\to kl,I}=1&+\frac{\alpha_{\rm s}}{\pi}\left({\cal C}^{\rm Coul,(1)}_{ij\to kl,I}(N,\{m^2\},\mu^2)+{\cal C}^{\rm (1)}_{ij\to kl,I}(\{m^2\},\mu^2)\right)\nonumber\\
&+\frac{\alpha_{\rm s}^2}{\pi^2}\left({\cal C}^{\rm Coul,(2)}_{ij\to kl,I}(N,\{m^2\},\mu^2)+{\cal C}^{\rm (2)}_{ij\to kl,I}(\{m^2\},\mu^2)\right. \nonumber\\
&\quad\quad\quad +\left. {\cal C}^{\rm (1)}_{ij\to kl,I}(\{m^2\},\mu^2){\cal C}^{\rm Coul,(1)}_{ij\to kl,I}(N,\{m^2\},\mu^2)\right).\label{eq:matchingcoeff}
\end{align} 
The first-order hard matching coefficients ${\cal C}^{\rm (1)}_{ij\to kl,I}$ were calculated in~\cite{Beenakker:2013mva}, whereas the expressions for the first-order Coulomb corrections ${\cal C}^{\rm Coul,(1)}_{ij\to kl,I}$ in Mellin-moment space are listed in appendix~\ref{app:1lcoulomb}. The form of the two-loop Coulomb corrections in $\beta$-space is known in the literature~\cite{Beneke:2009ye}. We calculate the ${\cal C}^{\rm Coul,(2)}_{ij\to kl,I}$ coefficient by taking Mellin moments of the near-threshold approximation of these two-loop Coulomb corrections, the result of which can be found in appendix~\ref{app:2lcoulomb}.  The second-order hard matching coefficient ${\cal C}^{\rm (2)}_{ij\to kl,I}$ is not known at the moment and we put  ${\cal C}^{\rm (2)}_{ij\to kl,I}=0$ in \eqref{eq:matchingcoeff}.

Once we have the NNLL resummed cross section in Mellin-moment space, we match it to the approximated NNLO cross section, which is constructed by adding the near-threshold approximation of the NNLO correction~\cite{Beneke:2009ye} to the full NLO result~\cite{Beenakker:1996ch}. The matching is performed according to 
\begin{align}
  \label{eq:matching}
  &\sigma^{\rm (NNLL~matched)}_{h_1 h_2 \to kl}\bigl(\rho, \{m^2\},\mu^2\bigr) 
  =\; \sigma^{\rm (NNLO_{Approx})}_{h_1 h_2 \to kl}\bigl(\rho, \{m^2\},\mu^2\bigr)
          \\[1mm]
& +\, \sum_{i,j}\,\int_\mathrm{CT}\,\frac{dN}{2\pi i}\,\rho^{-N}\,
       \tilde f_{i/h_1}(N+1,\mu^2)\,\tilde f_{j/h_{2}}(N+1,\mu^2) \nonumber\\[2mm]
&\times\,
       \left[\tilde\sigma^{\rm(res,NNLL)}_{ij\to kl}\bigl(N,\{m^2\},\mu^2\bigr)
             \,-\, \tilde\sigma^{\rm(res,NNLL)}_{ij\to kl}\bigl(N,\{m^2\},\mu^2\bigr)
       {\left.\right|}_{\scriptscriptstyle{\rm (NNLO_{Approx})}}\, \right]. \nonumber
\end{align}
To evaluate the inverse Mellin transform in~\eqref{eq:matching} we adopt the ``minimal prescription'' of reference~\cite{Catani:1996yz} for the integration contour CT. We will refer to the second term, i.e. $$\sum_{i,j}\,\int_\mathrm{CT}\,\frac{dN}{2\pi i}\,\rho^{-N}\,\tilde f_{i/h_1}(N+1,\mu^2)\,\tilde f_{j/h_{2}}(N+1,\mu^2) \tilde\sigma^{\rm(res,NNLL)}_{ij\to kl}\bigl(N,\{m^2\},\mu^2\bigr)$$ as the resummed part of the cross section, while the third term $$\sum_{i,j}\,\int_\mathrm{CT}\,\frac{dN}{2\pi i}\,\rho^{-N}\,\tilde f_{i/h_1}(N+1,\mu^2)\,\tilde f_{j/h_{2}}(N+1,\mu^2) \tilde\sigma^{\rm(res,NNLL)}_{ij\to kl}\bigl(N,\{m^2\},\mu^2\bigr) {\left.\right|}_{\scriptscriptstyle{\rm (NNLO_{Approx})}}$$ provides the NNLL resummed cross section expanded to NNLO accuracy.

%%%%%%%%%%%%%%%%%%%%%%%%%%%%%%%%%%%%%%%%

\section{Numerical results}

%%%%%%%%%%%%%%%%%%%%%%%%%%%%%%%%%%%%%%%%

\label{s:numerics}

In this section we present numerical results for the NNLL resummed cross sections matched to the approximated NNLO results for pair production of squarks and gluinos at the LHC with $\sqrt S = 8$ TeV.

All flavours of final-state squarks are included and summed over, except top squarks, due to the large mixing effects and the mass splitting in the stop sector~\cite{Ellis:1983ed}. We sum over squarks with both chiralities ($\tilde{q}_{L}$ and~$\tilde{q}_{R}$), which are taken as mass degenerate. All light-flavour squarks are also assumed to be mass degenerate. The QCD coupling $\alpha_{\rm s}$ and the parton distribution functions at NLO and NNLO are defined in the $\overline{\rm MS}$ scheme with five active flavours. The renormalisation and factorisation scales are taken to be equal $\mu=\mu_R=\mu_F$ and a top-quark mass of $m_t=173.07$~GeV~\cite{Beringer:1900zz} is used.

As our default for NNLL and approximated NNLO calculations, we use the MSTW 2008 NNLO parton distribution function (pdfs)~\cite{Martin:2009iq} with the corresponding $\as(M_Z)=0.117$. The NLO and NLL results presented for reference are obtained using the MSTW 2008 NLO parton distribution functions~\cite{Martin:2009iq} with the corresponding $\alpha_{\rm s}(M_{Z}) = 0.120$. In order to use standard parametrisations of pdfs in $x$-space we employ the method introduced in reference~\cite{Kulesza:2002rh}. Alternatively, where appropriate, we use the program {\tt PEGASUS}~\cite{Vogt:2004ns} to derive the Mellin moments of the pdfs based on the MSTW parametrisation at the initial factorisation scale~\cite{Martin:2009iq}.

In the following discussion we present predictions for the LHC squark and gluino cross sections for a center-of-mass energy of 8~TeV,  at various levels of theoretical accuracy:

\begin{itemize}

\item The NLO cross sections~\cite{Beenakker:1996ch}, denoted as $\sigma^{\rm NLO}$.

\item The NLL cross sections matched to NLO results, based on the calculations presented in \cite{Beenakker:2009ha,Kulesza:2008jb,Kulesza:2009kq} and using the MSTW 2008 NLO parton distribution functions. They are denoted as $\sigma^{\rm NLO+NLL}$.

\item The approximated NNLO cross sections, calculated by adding near-threshold approximations of the NNLO corrections~\cite{Beneke:2009ye} to the NLO cross sections. These cross sections are denoted as $\sigma^{\rm NNLO_{Approx}}$.

\item The NNLL matched cross sections $\sigma^{\rm NNLL~matched}$. The NNLO$_{\rm Approx}$ + NNLL accuracy, as detailed in Eq.~(\ref{eq:matching}), applies to the $s$-wave channels. The contributions from the $\beta^2$-suppressed $p$-wave channels are taken into account at NLO+NLL accuracy. Both the $s$-wave and the $p$-wave contributions have been convoluted with the MSTW 2008 NNLO parton distribution functions. We have checked that using NNLO pdfs instead of NLO pdfs for the suppressed $p$-wave channel contributions leads to a negligible modification of the full result.

\end{itemize}
Apart from the NLO cross sections, which were calculated using the publicly available {\tt PROSPINO} code~\cite{prospino}, all our results were obtained using two independent computer codes. We have chosen squark and gluino masses going up to 2.5 TeV, as with the reported integrated luminosities by the ATLAS and CMS experiments, a few events are observable in the mass range from 2 to 2.5 TeV for the cross section on inclusive coloured sparticle production. In addition, the behaviour of the cross sections at large masses is interesting from a theoretical point of view, and the choice of the mass range furthermore leads to an easy comparison with the results obtained in soft-collinear effective theory.
We now discuss our findings.

\begin{figure*}

\begin{tabular}{ll}

(a)\includegraphics[width=0.45\columnwidth]{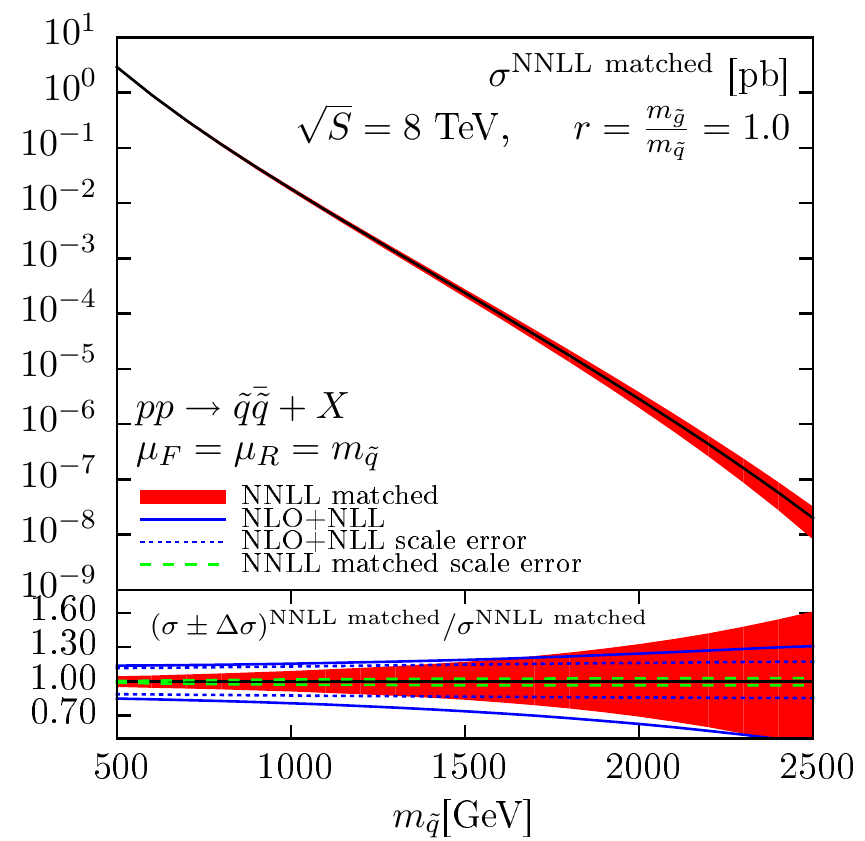}&

(b)\includegraphics[width=0.45\columnwidth]{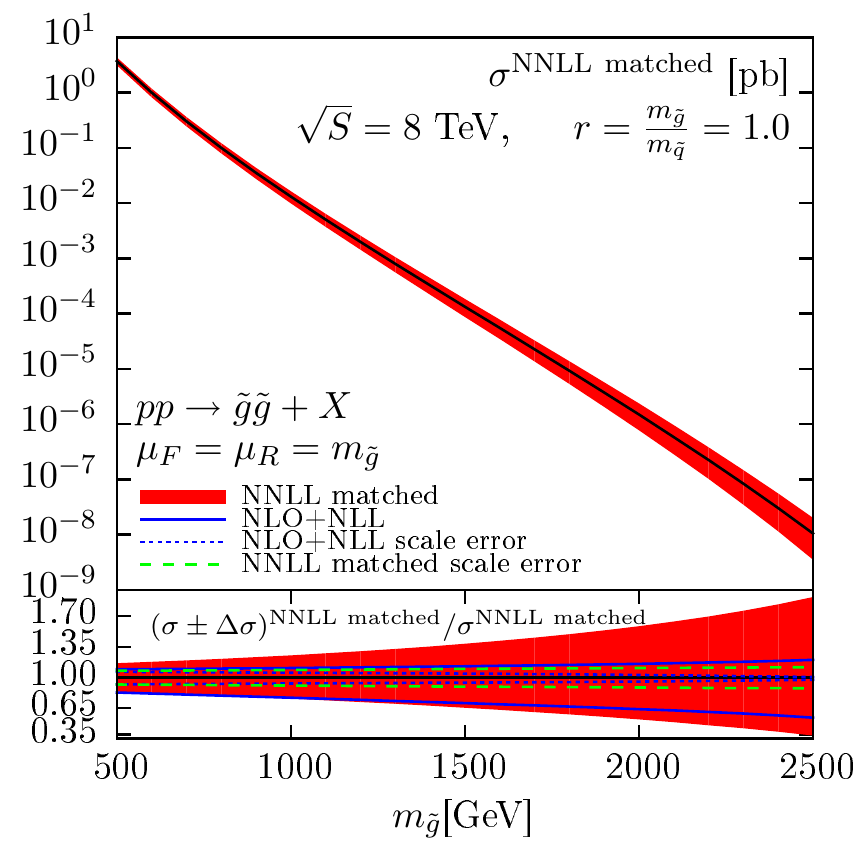}\\

(c)\includegraphics[width=0.45\columnwidth]{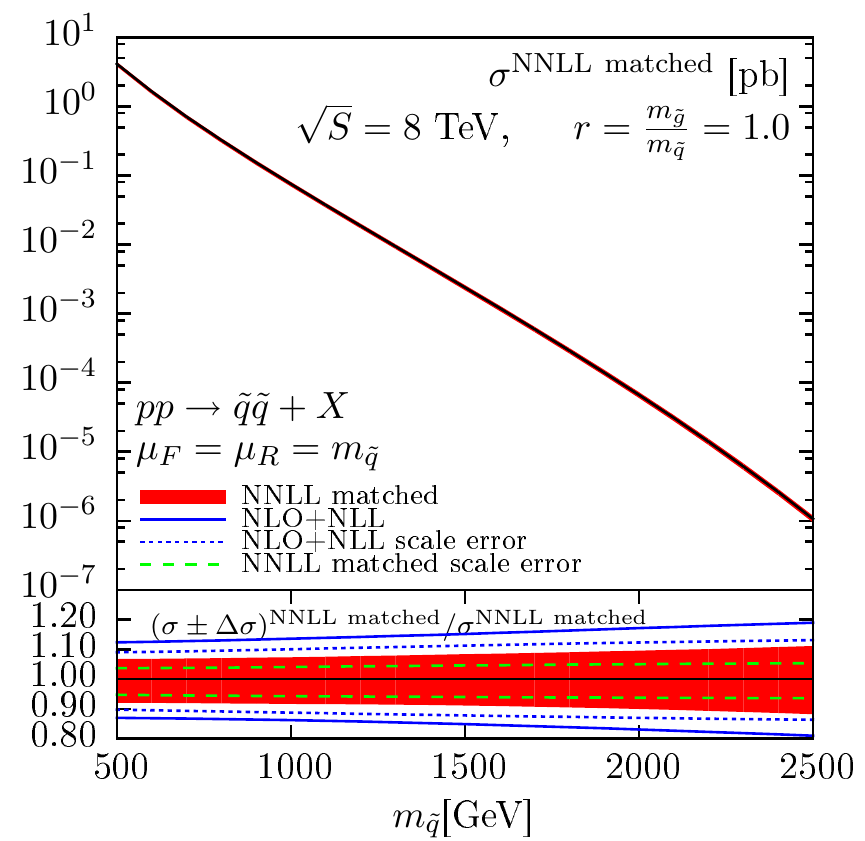}&

(d)\includegraphics[width=0.45\columnwidth]{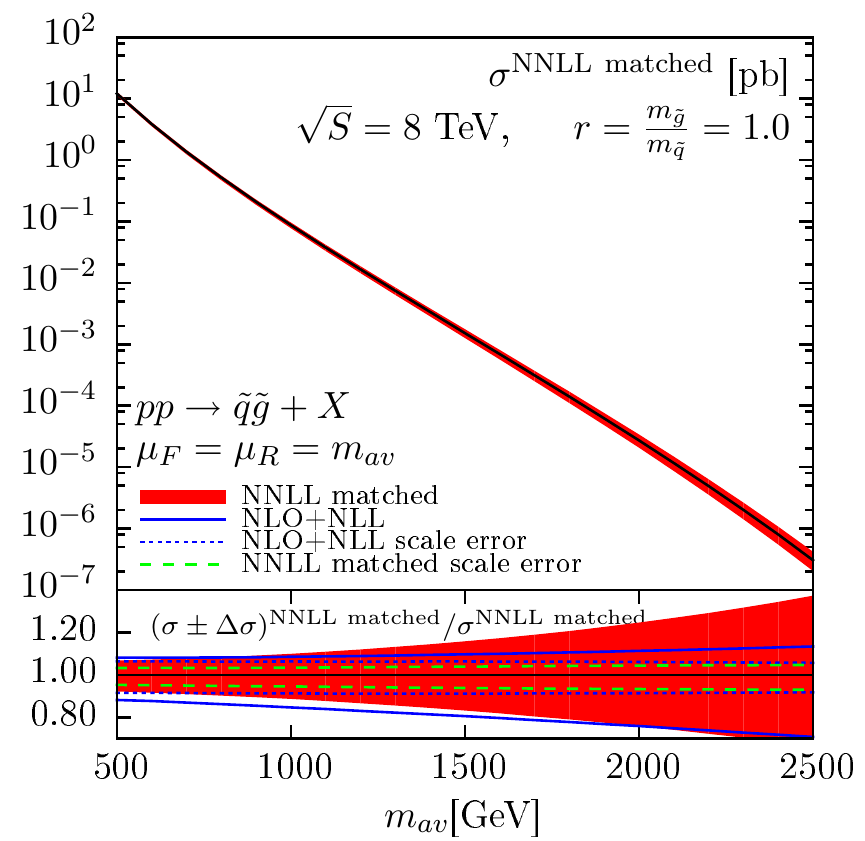}

\end{tabular}\caption{The NNLL matched cross section for the four processes of pair production of squarks and gluinos, including the theoretical error band for the NNLO approximation. The error band includes the 68\% C.L. pdf and $\alpha_{\rm s}$ errors, added quadratically, and the scale uncertainty varied in the range $m_{av}/2 \leq \mu \leq 2m_{av}$, added linearly to the combined pdf and $\alpha_{\rm s}$ error. The energy is that of the LHC at 8 TeV. The squark and gluino masses have been taken equal and the common renormalisation and factorisation scale has been set equal to the average mass of the two particles produced. For comparison, we also show the scale and total uncertainty of the NLO+NLL prediction.\label{fig:error}}

\end{figure*}

\begin{figure*}

\centering

\includegraphics[width=0.45\columnwidth]{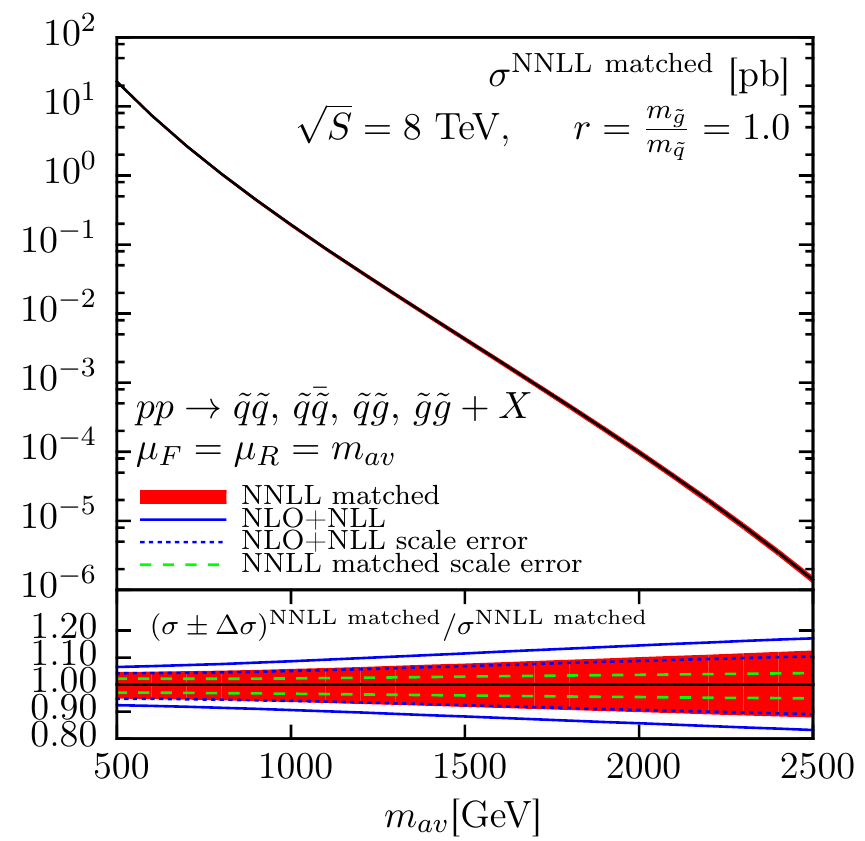}

\caption{The NNLL matched cross section for the sum of the four processes of pair production of squarks and gluinos, including the theoretical error band for the NNLO approximation. The error band includes the 68\% C.L. pdf and $\alpha_{\rm s}$ errors, added quadratically, and the scale uncertainty varied in the range  $m_{av}/2 \leq \mu \leq 2m_{av}$, added linearly to the combined pdf and $\alpha_{\rm s}$ error. The energy is that of the LHC at 8 TeV. The squark and gluino masses have been taken equal and the common renormalisation and factorisation scale has been set equal to the average mass of the two particles produced. For comparison, we also show the scale and total uncertainty of the NLO+NLL prediction. \label{fig:susy-error}}

\end{figure*}

\begin{figure*}

		\begin{tabular}{ll}

			(a)\includegraphics[width=0.45\columnwidth]{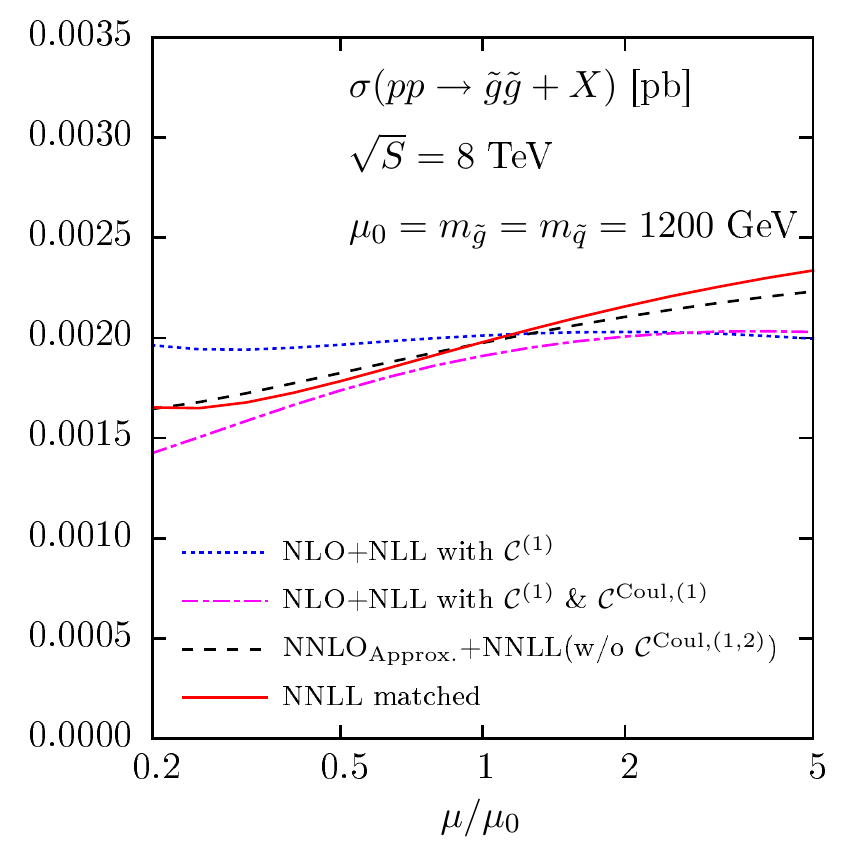}&

			(b)\includegraphics[width=0.45\columnwidth]{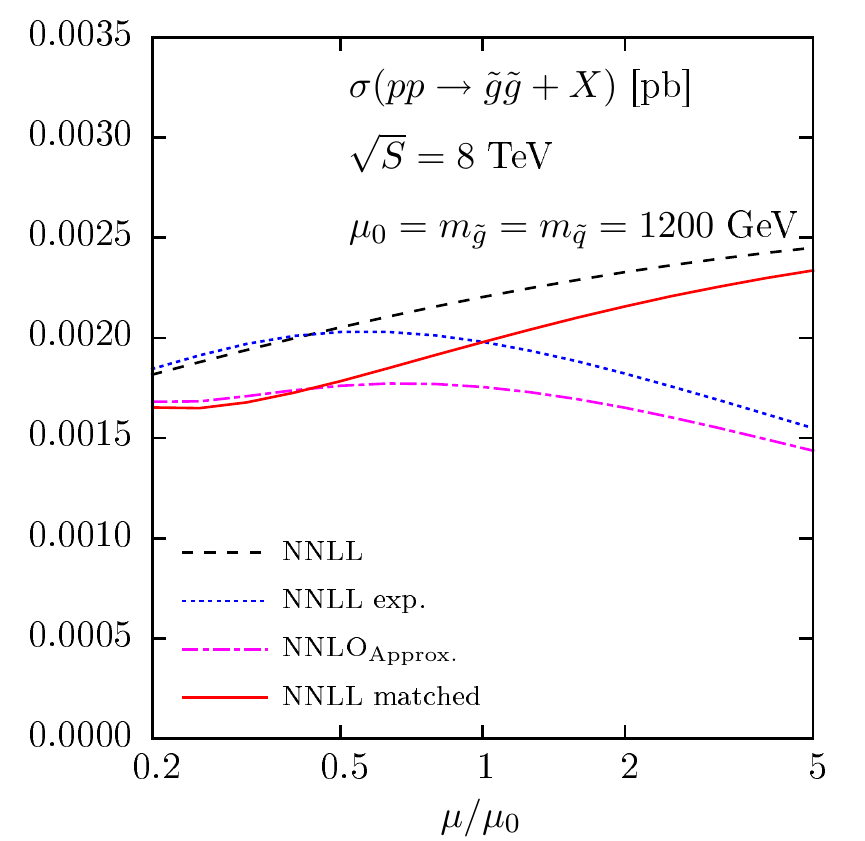}

		\end{tabular}

		\caption{Scale dependence of different parts of the gluino pair production cross section: intermediate levels of accuracy (a), and a split-up of the cross section into resummed and expanded parts for NNLL matched, compared to the full NNLL matched result as well as $\mathrm{NNLO_{Approx.}}$ (b).\label{fig:intmscale}}

	\end{figure*}

In figures~\ref{fig:error} and \ref{fig:susy-error}, and in table~\ref{tab:nnll} we present the NNLL matched cross section predictions for the four different pair-production processes for squarks and gluinos as well as the sum of the NNLL matched cross sections for these four SUSY-QCD processes. The theoretical uncertainty includes the scale error as well as pdf and $\alpha_{\rm s}$ errors. It is obtained by linearly adding the scale dependence in the range $m/2 \leq \mu \leq 2m$ to the combined 68\% C.L. pdf and $\alpha_{\rm s}$ uncertainties, the latter two added in quadrature.
We see uncertainties grow from small for masses near the present lower bounds, to sizeable for masses approaching 2.5 TeV. For comparison, we also show the NLO+NLL scale and total uncertainty in figures~\ref{fig:error} and \ref{fig:susy-error}. We find that the scale uncertainty at NNLL is reduced with respect to the NLO+NLL prediction, except for gluino-pair production. We return to the scale dependence of the NNLL gluino-pair cross section later in this section, see figure~\ref{fig:intmscale} and the corresponding text. On the other hand, the pdf and $\alpha_{\rm s}$ errors are larger for the NNLL matched than for the NLO+NLL prediction, and dominate the total NNLL matched uncertainty, in particular at larger squark and gluino masses. We note that this increase is driven by the increase in the $\alpha_{\rm s}$ uncertainty when going from the NLO to NNLO MSTW pdfs. See table~\ref{tab:nnll} for a breakdown of the uncertainties of the NNLL matched prediction.

\renewcommand{\arraystretch}{1.3}

\begin{table}
\begin{centering}
\begin{tabular}{|c|c|c|c|c|c|c|}
\hline
process & $m$[GeV] & $\sigma^{\rm NNLL~matched}$[pb] & Scale Error & pdf Error & $\alpha_{\rm s}$~Error & Total~Error \\
\hline
$\sq\sqb$ & $500$ & $2.90$ & $~^{+0.37\%}_{-1.2\%}$ & $~^{+2.6\%}_{-2.3\%}$ & $~^{+3.2\%}_{-2.8\%}$ & $~^{+4.5\%}_{-4.8\%}$ \\
$\sq\sqb$ & $1000$ & $1.80\times 10^{-2}$ & $~^{+1.7\%}_{-2.8\%}$ & $~^{+6.0\%}_{-5.3\%}$ & $~^{+4.5\%}_{-2.9\%}$ & $~^{+9.2\%}_{-8.8\%}$ \\
$\sq\sqb$ & $1500$ & $2.37\times 10^{-4}$ & $~^{+2.4\%}_{-3.3\%}$ & $~^{+11\%}_{-12\%}$ & $~^{+9.6\%}_{-5.2\%}$ & $~^{+17\%}_{-16\%}$ \\
$\sq\sqb$ & $2000$ & $2.86\times 10^{-6}$ & $~^{+2.7\%}_{-3.3\%}$ & $~^{+20\%}_{-25\%}$ & $~^{+22\%}_{-12\%}$ & $~^{+33\%}_{-31\%}$ \\
$\sq\sqb$ & $2500$ & $2.01\times 10^{-8}$ & $~^{+3.0\%}_{-3.3\%}$ & $~^{+37\%}_{-50\%}$ & $~^{+45\%}_{-24\%}$ & $~^{+61\%}_{-59\%}$ \\
\hline
$\gl\gl$ & $500$ & $3.72$ & $~^{-8.1\%}_{+7.6\%}$ & $~^{+6.7\%}_{-6.8\%}$ & $~^{+5.5\%}_{-4.3\%}$ & $~^{+16\%}_{-16\%}$ \\
$\gl\gl$ & $1000$ & $1.30\times 10^{-2}$ & $~^{-9.3\%}_{+8.6\%}$ & $~^{+12\%}_{-13\%}$ & $~^{+11\%}_{-7.3\%}$ & $~^{+25\%}_{-24\%}$ \\
$\gl\gl$ & $1500$ & $1.33\times 10^{-4}$ & $~^{-10\%}_{+9.7\%}$ & $~^{+20\%}_{-21\%}$ & $~^{+21\%}_{-12\%}$ & $~^{+38\%}_{-35\%}$ \\
$\gl\gl$ & $2000$ & $1.49\times 10^{-6}$ & $~^{-12\%}_{+11\%}$ & $~^{+31\%}_{-31\%}$ & $~^{+36\%}_{-19\%}$ & $~^{+59\%}_{-48\%}$ \\
$\gl\gl$ & $2500$ & $1.05\times 10^{-8}$ & $~^{-12\%}_{+12\%}$ & $~^{+49\%}_{-45\%}$ & $~^{+64\%}_{-30\%}$ & $~^{+92\%}_{-66\%}$ \\
\hline
$\sq\sq$ & $500$ & $4.13$ & $~^{+3.7\%}_{-5.2\%}$ & $~^{+2.6\%}_{-2.0\%}$ & $~^{+1.7\%}_{-1.7\%}$ & $~^{+6.8\%}_{-7.9\%}$ \\
$\sq\sq$ & $1000$ & $7.51\times 10^{-2}$ & $~^{+4.1\%}_{-5.7\%}$ & $~^{+3.3\%}_{-2.5\%}$ & $~^{+0.41\%}_{-0.50\%}$ & $~^{+7.4\%}_{-8.3\%}$ \\
$\sq\sq$ & $1500$ & $2.39\times 10^{-3}$ & $~^{+4.6\%}_{-6.0\%}$ & $~^{+3.7\%}_{-2.8\%}$ & $~^{-0.60\%}_{+0.48\%}$ & $~^{+8.4\%}_{-8.9\%}$ \\
$\sq\sq$ & $2000$ & $6.66\times 10^{-5}$ & $~^{+5.0\%}_{-6.2\%}$ & $~^{+4.3\%}_{-3.5\%}$ & $~^{-1.5\%}_{+1.5\%}$ & $~^{+9.6\%}_{-10\%}$ \\
$\sq\sq$ & $2500$ & $1.07\times 10^{-6}$ & $~^{+5.4\%}_{-6.4\%}$ & $~^{+5.2\%}_{-4.7\%}$ & $~^{-2.5\%}_{+2.5\%}$ & $~^{+11\%}_{-12\%}$ \\
\hline
$\sq\gl$ & $500$ & $12.1$ & $~^{-4.7\%}_{+3.2\%}$ & $~^{+2.3\%}_{-2.1\%}$ & $~^{+2.7\%}_{-2.3\%}$ & $~^{+6.7\%}_{-7.8\%}$ \\
$\sq\gl$ & $1000$ & $8.74\times 10^{-2}$ & $~^{-5.4\%}_{+3.4\%}$ & $~^{+4.9\%}_{-5.0\%}$ & $~^{+4.4\%}_{-3.1\%}$ & $~^{+10\%}_{-11\%}$ \\
$\sq\gl$ & $1500$ & $1.52\times 10^{-3}$ & $~^{-6.1\%}_{+3.9\%}$ & $~^{+8.6\%}_{-9.3\%}$ & $~^{+8.2\%}_{-5.1\%}$ & $~^{+16\%}_{-17\%}$ \\
$\sq\gl$ & $2000$ & $2.72\times 10^{-5}$ & $~^{-6.7\%}_{+4.4\%}$ & $~^{+14\%}_{-15\%}$ & $~^{+15\%}_{-8.7\%}$ & $~^{+25\%}_{-24\%}$ \\
$\sq\gl$ & $2500$ & $3.05\times 10^{-7}$ & $~^{-7.0\%}_{+4.7\%}$ & $~^{+21\%}_{-22\%}$ & $~^{+24\%}_{-14\%}$ & $~^{+37\%}_{-33\%}$ \\
\hline
inclusive & $500$ & $22.9$ & $~^{+2.2\%}_{-3.0\%}$ & $~^{+1.7\%}_{-1.6\%}$ & $~^{+1.8\%}_{-1.5\%}$ & $~^{+4.7\%}_{-5.2\%}$ \\
inclusive & $1000$ & $1.94\times 10^{-1}$ & $~^{+2.3\%}_{-3.4\%}$ & $~^{+2.7\%}_{-2.7\%}$ & $~^{+2.2\%}_{-1.5\%}$ & $~^{+5.8\%}_{-6.4\%}$ \\
inclusive & $1500$ & $4.29\times 10^{-3}$ & $~^{+2.9\%}_{-4.0\%}$ & $~^{+3.8\%}_{-3.8\%}$ & $~^{+3.0\%}_{-1.9\%}$ & $~^{+7.8\%}_{-8.2\%}$ \\
inclusive & $2000$ & $9.81\times 10^{-5}$ & $~^{+3.6\%}_{-4.6\%}$ & $~^{+4.9\%}_{-4.9\%}$ & $~^{+4.3\%}_{-2.7\%}$ & $~^{+10\%}_{-10\%}$ \\
inclusive & $2500$ & $1.40\times 10^{-6}$ & $~^{+4.2\%}_{-5.1\%}$ & $~^{+6.2\%}_{-6.1\%}$ & $~^{+5.7\%}_{-3.6\%}$ & $~^{+13\%}_{-12\%}$ \\
\hline
\end{tabular}
\caption{The NNLL matched cross section prediction for different SUSY-QCD processes at the LHC with $\sqrt S=8$ TeV. The scale, pdf, $\alpha_{\rm s}$ and total uncertainty are shown separately. The MSTW 2008 NNLO pdfs~\cite{Martin:2009iq} have been adopted, the squark and gluino masses have been taken equal and the common renormalisation and factorisation scale has been set equal to the squark/gluino mass.\label{tab:nnll}}
\end{centering}
\end{table}

\begin{figure*}

\hspace{-0.55cm}

\begin{tabular}{ll}

(a)\includegraphics[width=0.45\columnwidth]{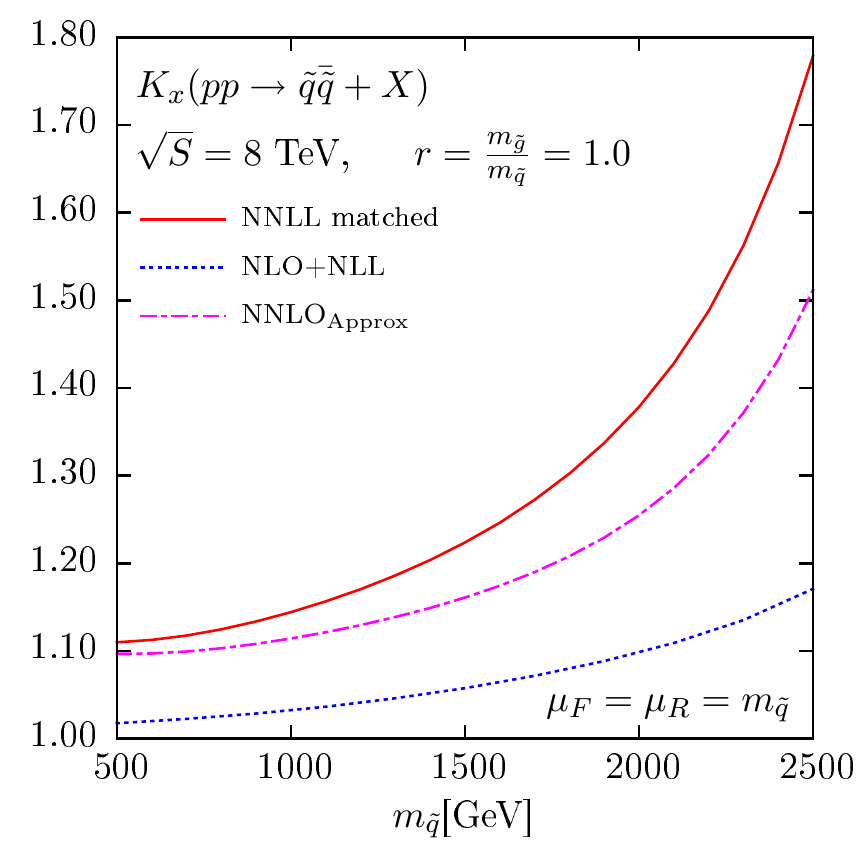}&

(b)\includegraphics[width=0.45\columnwidth]{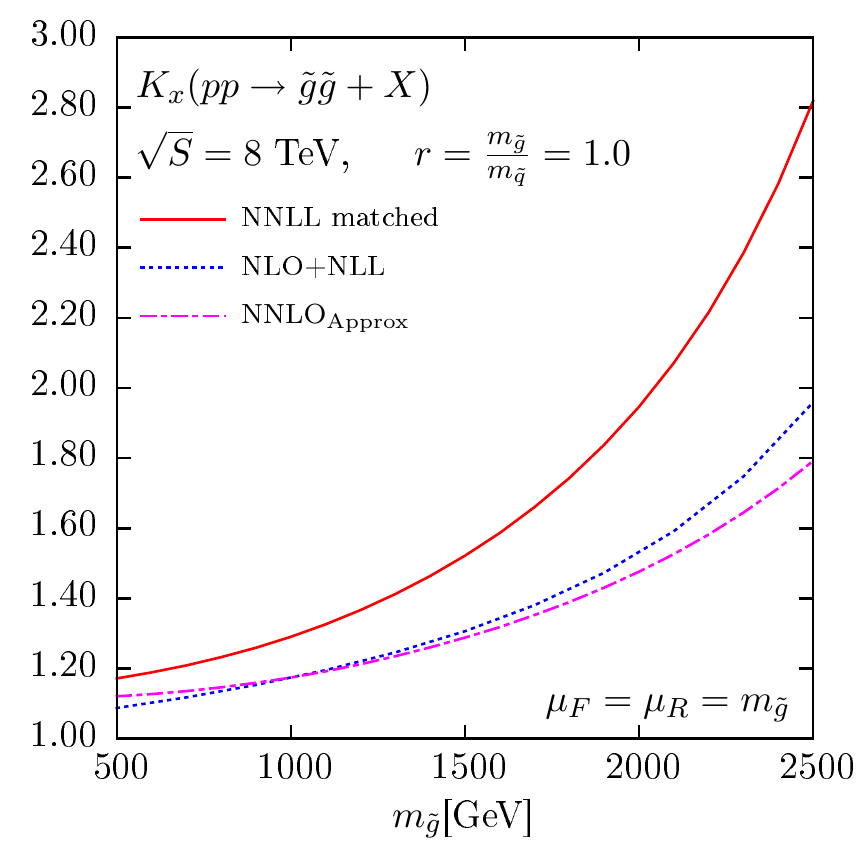}\\

(c)\includegraphics[width=0.45\columnwidth]{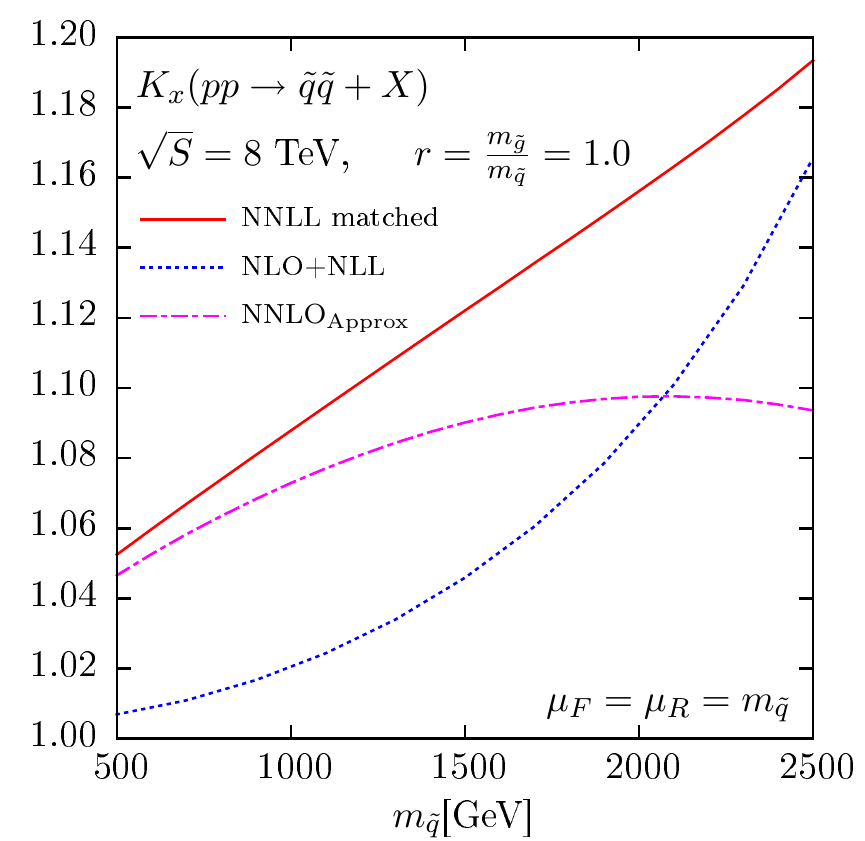}&

(d)\includegraphics[width=0.45\columnwidth]{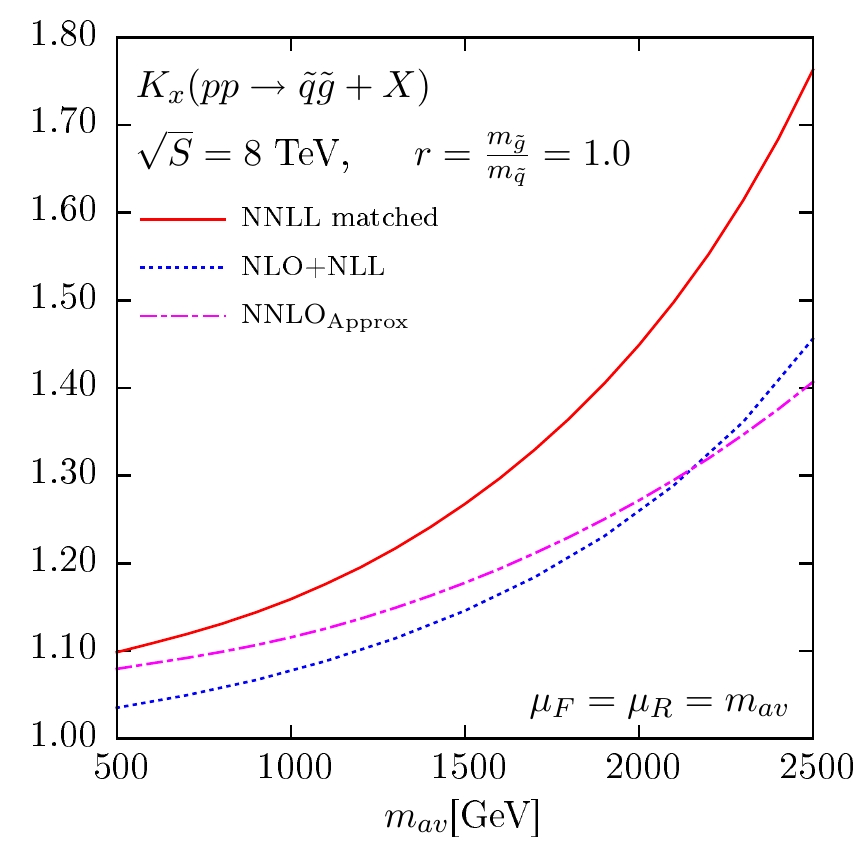}

\end{tabular}\caption{The $K_X$-factor for the NNLL matched, NLO+NLL and $\rm NNLO_{Approx}$ predictions for different SUSY-QCD processes at the LHC with $\sqrt S=8$ TeV. The squark and gluino masses have been taken equal and the common renormalisation and factorisation scale has been set equal  to the average mass of the two particles produced.\label{fig:K}}

\end{figure*}

Next we study how large the various corrections are with respect to the NLO corrections. For this purpose we define the $K$-factor:
\[K_{X}=\frac{\sigma^{X}}{\sigma^{\rm NLO}}~,\]
where $X$ indicates the accuracy of the considered predictions. The mass dependence of the $K_X$-factors for the four squark and gluino production processes in the case of equal squark and gluino masses is shown in figure~\ref{fig:K}. 

The NNLL corrections are positive and become larger as the masses of the produced particles increase. This is to be expected, since in general higher masses cause a process to take place closer to the production threshold. For all processes, the NNLL corrections lead to higher cross sections than previously known for the NLO+NLL results. The largest effect can be observed for gluino-pair production with the NNLL $K_{NNLL}$-factor approaching a value of 3 for very high gluino masses.

Compared to the other processes, we observe for $\sq\sq$ production a less rapid increase of the cross section due to NNLL corrections. There are two effects playing a role. The first effect is caused by the one-loop Coulomb corrections. The one-loop Coulomb correction for the sextet colour structure is negative, leading to a reduction of the cross section, whereas the opposite holds for the antitriplet colour structure. The importance of the negative sextet contribution is enhanced due to the change in the relative size of the LO contributions occurring with increasing mass, i.e.~the sextet colour channel becoming the dominant one at high mass, as opposed to the antitriplet colour channel being dominant at low mass. For comparison, in $\sq\gl$ production the negative contribution due to the Coulomb correction for the 15-plet colour channel is not as dominant when compared to contributions from other colour channels. Additionally the overall contribution of the Coulomb terms compared to the logarithic terms is smaller for $\sq\gl$ than for $\sq\sq$. The second effect is due to the difference between the behaviour of the NNLO pdfs and the NLO pdfs for valence quarks and gluons. The ratio of NNLO pdfs to NLO pdfs rises with an increasing value of $x$ for gluons, whereas it decreases for quarks. For all processes with gluons in the initial state this leads to an enhancement of the NNLL and NNLO corrections w.r.t. the NLO+NLL results that grows with increasing final-state masses. Conversely, for the processes with quarks in the initial state this leads to a suppression of these corrections at larger masses. The production of $\sq\sq$ pairs is the only process that at LO takes place exclusively in the $qq$ channel.

\begin{figure*}

\hspace{-0.55cm}

\begin{tabular}{ll}

(a)\includegraphics[width=0.45\columnwidth]{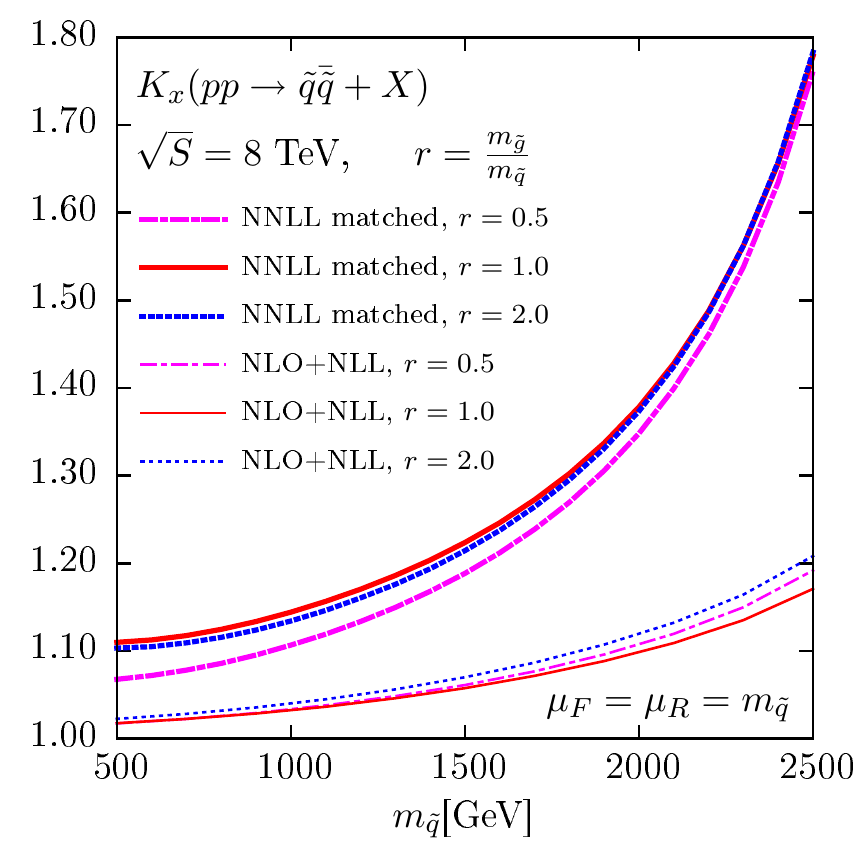}&

(b)\includegraphics[width=0.45\columnwidth]{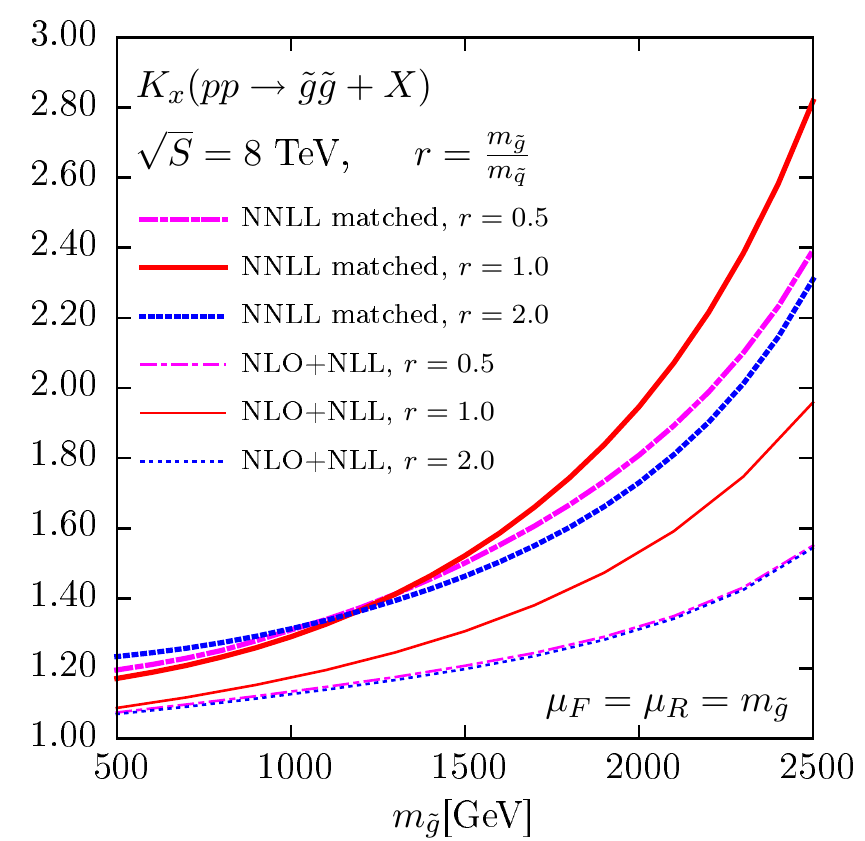}\\

(c)\includegraphics[width=0.45\columnwidth]{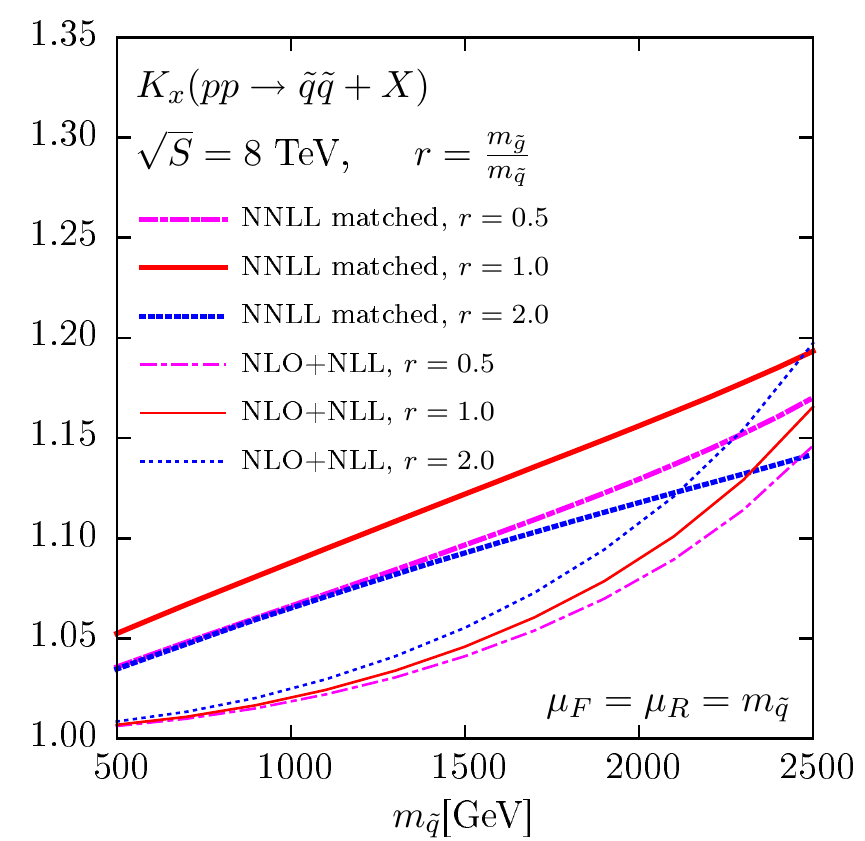}&

(d)\includegraphics[width=0.45\columnwidth]{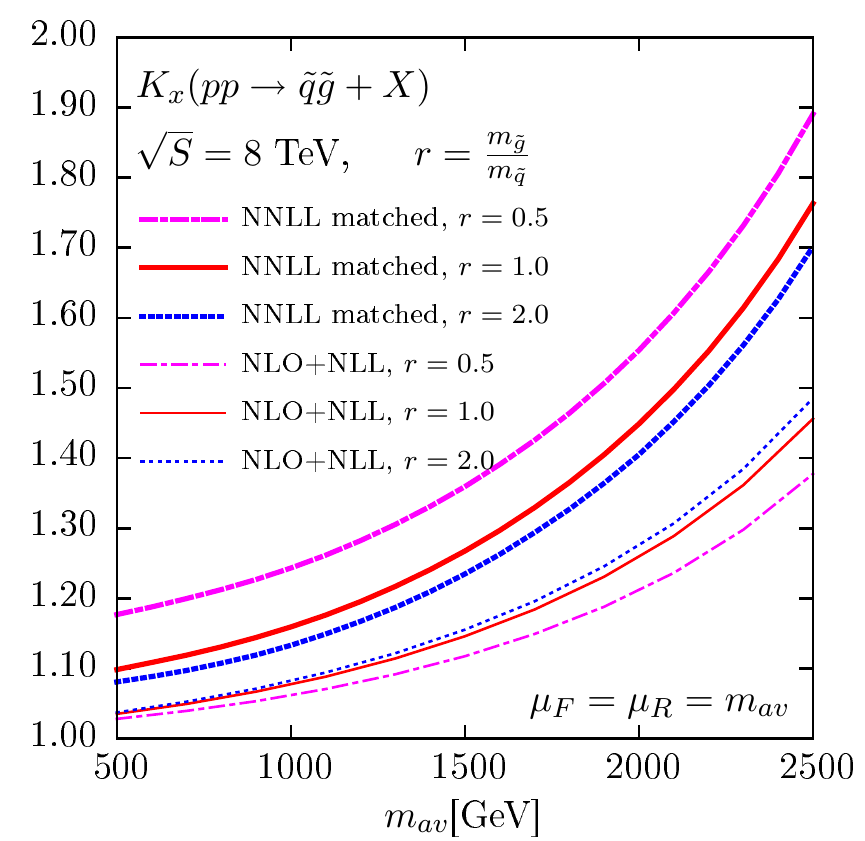}

\end{tabular}\caption{The $K_X$-factor for the NNLL matched and NLO+NLL predictions for different values of $r=m_{\gl}/m_{\sq}=0.5,1,2$ for the four pair-production processes of squarks and gluinos at the LHC with $\sqrt S=8$ TeV. The common renormalisation and factorisation scale has been set equal to the average mass of the two particles produced.\label{fig:K-r}}

\end{figure*}

In figure~\ref{fig:K-r} we show the mass dependence of the $K$-factor for different ratios of the squark and gluino masses. The differences in the corrections for various mass ratios can be largely explained by the $r$-dependence of the matching coefficients $C^{\rm NNLL}_{ij \to kl,I}$, composed of the hard matching coefficients~\cite{Beenakker:2013mva} and the Coulomb corrections. Moreover, the two-loop Coulomb coefficients depend on $r$ only for $\tilde{q}\tilde{g}$ production, causing the additional $r$-dependence of the $K_{NNLL}$-factor for this production process.

\begin{figure*}

\hspace{-0.55cm}

\begin{tabular}{ll}

(a)\includegraphics[width=0.45\columnwidth]{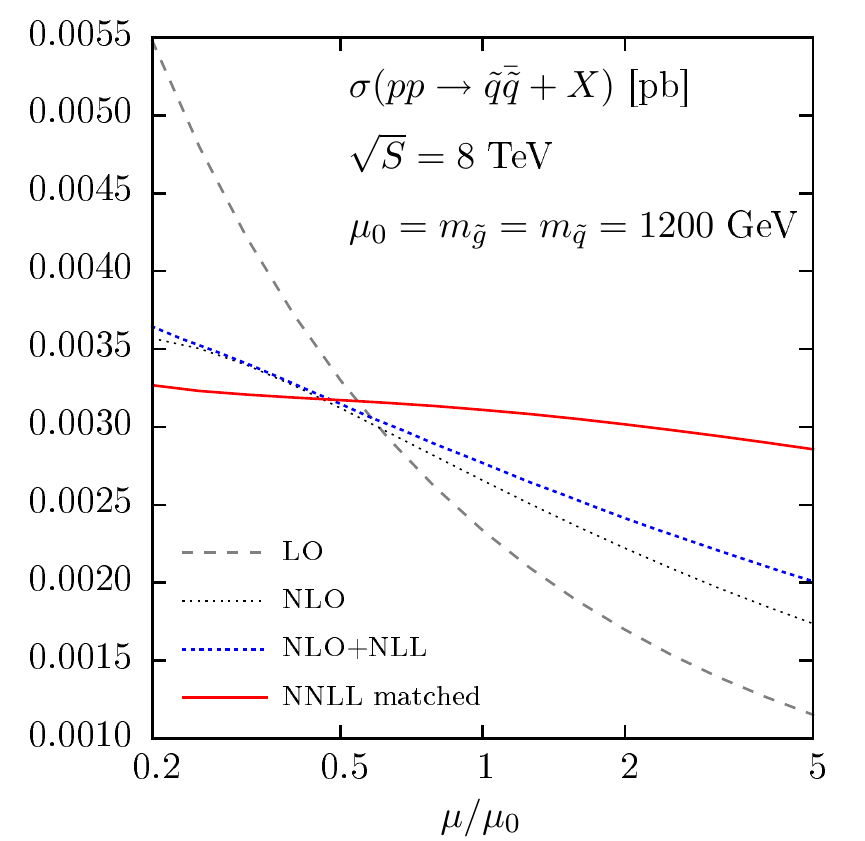}&

(b)\includegraphics[width=0.45\columnwidth]{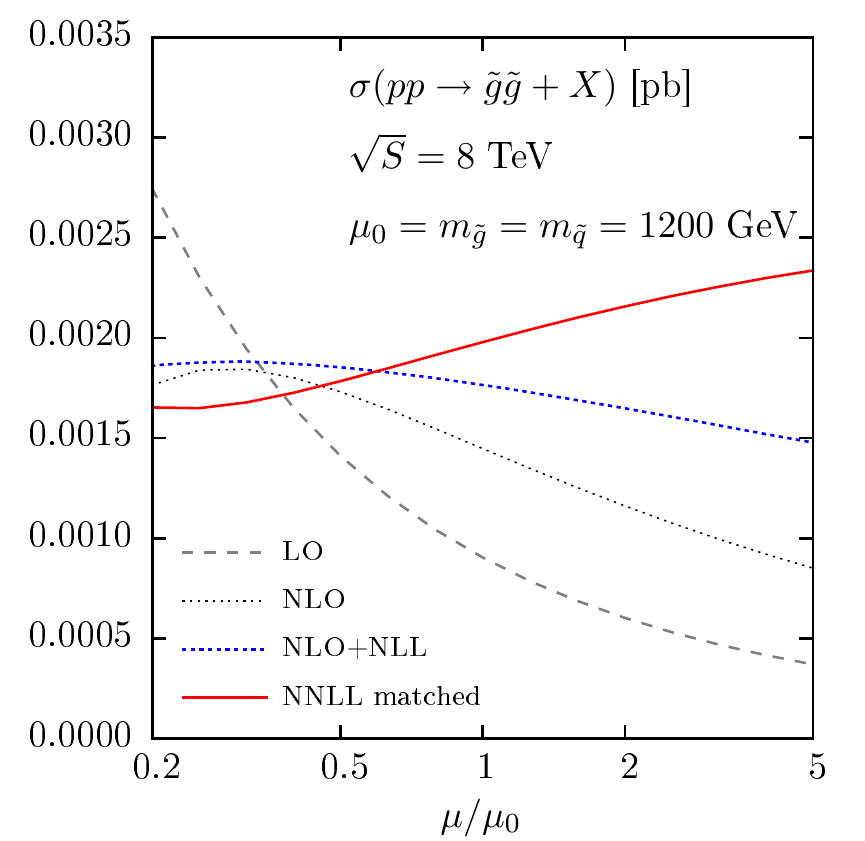}\\

(c)\includegraphics[width=0.45\columnwidth]{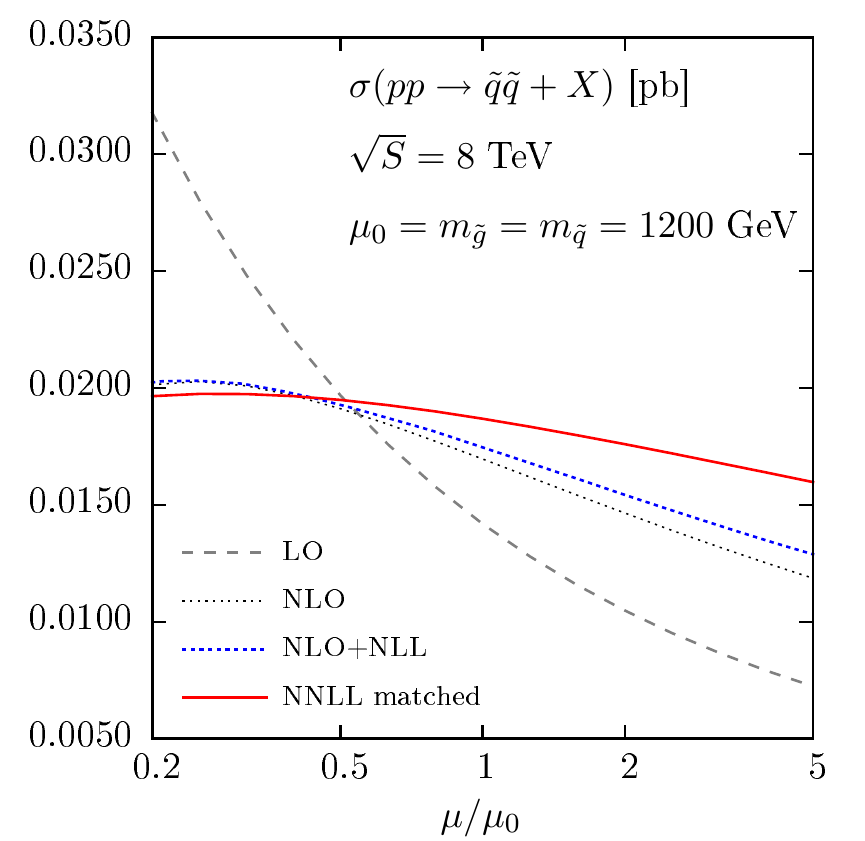}&

(d)\includegraphics[width=0.45\columnwidth]{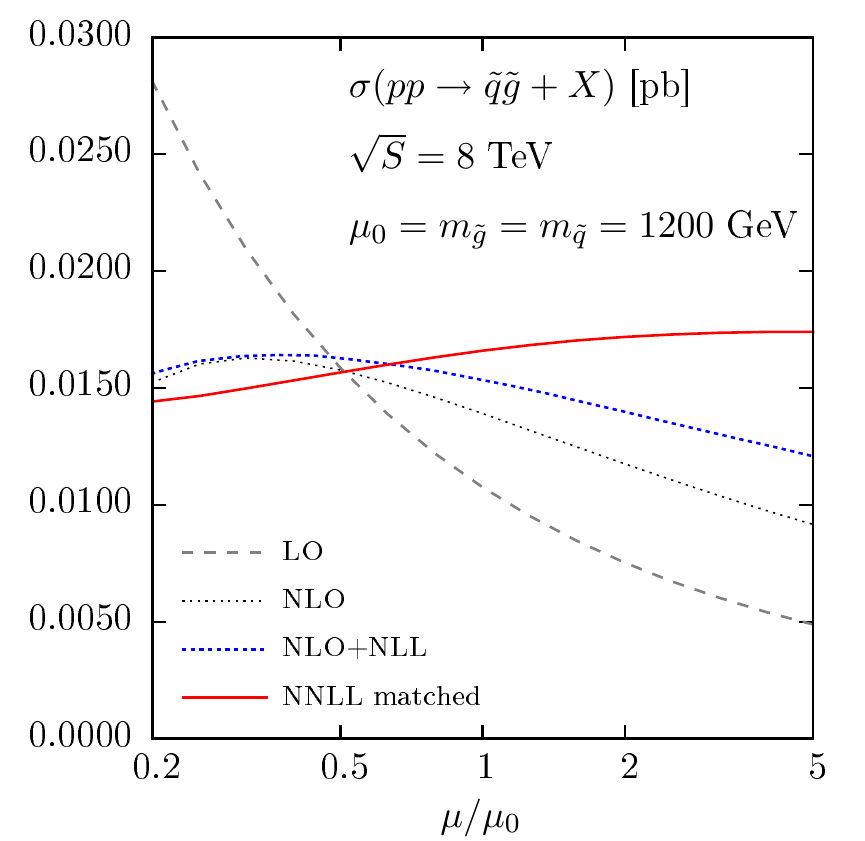}

\end{tabular}\caption{Scale dependence of the LO, NLO, NLO+NLL and NNLL matched cross sections for the four different SUSY-QCD processes at the LHC with $\sqrt S=8$ TeV. The squark and gluino masses have been taken equal to 1.2 TeV.\label{fig:scale}}

\end{figure*}

In the next step we investigate the scale dependence for the different processes, see  figure~\ref{fig:scale}. For this analysis, the squark and gluino mass are taken to be equal to 1.2~TeV. We vary the scale around the central value of $\mu_0=m_{\sq}=m_{\gl}$ from $\mu=\mu_0/5$ to $\mu=5\mu_0$. We observe that the scale dependence decreases when including NNLL matched corrections for all processes except gluino-pair production.

	In order to understand the increased scale dependence of gluino-pair production, we analyse, apart from the NLO+NLL results and NNLL matched results, also the effects of incorporating specific terms of NNLL accuracy in the resummed cross sections. In addition to NNLL matched results, figure \ref{fig:intmscale} (a) shows the ``NLO+NLL with ${\cal C}^{(1)}$'' predictions, where the exponential in the resummed part is considered at NLL accuracy, but the formally NNLL hard matching coefficients ${\cal C}^{(1)}_{ij \to kl, I}$ are taken into account while all Coulomb corrections are kept equal to zero.  Furthermore, we present the ``NLO+NLL with ${\cal C}^{(1)}$ \& ${\cal C}^{\rm Coul,(1)}$'' results, which additionally include the one-loop Coulomb corrections ${\cal C}^{\rm Coul,(1)}_{ij \to kl, I}$, as well as the ``$\mathrm{NNLO_{Approx.}}$+NNLL(w/o ${\cal C}^{\rm Coul,(1,2)} )$'' results, which include the NNLL exponential and the hard matching coefficients. With the inclusion of one-loop Coulomb corrections at the NLL level, we notice a significant change in scale dependence. This implies that one-loop Coulomb corrections are, at least partly, the cause for the different behaviour of the cross section when varying the scale. The impact of the one-loop Coulomb corrections on the scale dependence of the predictions appears to be smaller when considered together with the NNLL exponential. Additionally, the contributions due to two-loop Coulomb corrections present at NNLL largely balance the effect of terms involving one-loop Coulomb terms.

	An analysis that involved splitting up the cross sections into their colour channels has revealed that the inclusion of one-loop Coulomb coefficients mainly affects the scale behaviour of the fixed-order expansion of the resummed cross section,  whereas it hardly changes the scale behaviour of the purely resummed part. The interplay between the opposite-sign Coulomb contributions for the attractive and repulsive colour channels in the expanded part leads to a constant shift for higher scales, and a change in the slope of the curve for smaller scales. The almost constant shift in all the colour channels of the resummed part cannot compensate for this change in behaviour of the expanded part.

	We notice another change when including logarithmic terms up to NNLL in the exponentials. The growth in scale dependence of the resummed part at large scales is driven by these logarithms. We note that in comparison to the other processes, the logarithmic terms are more important for the production of gluinos due to the colour factors. The importance of the higher-order logarithms can be seen even more clearly when comparing the resummed part of the cross section to its expansion up to NNLO, cf. figure \ref{fig:intmscale} (b). Whereas the resummed part (NNLL) and its expansion (NNLL exp.) agree well for low scales, we notice a difference at high scales, resulting from logarithms of formally higher order in $\as$ that are not included in the expansion. We have also verified that this difference is much smaller for the $\sq\sqb$ process, in agreement with the expected higher relevance of the logarithmic terms for gluino-pair production due to more intense gluon radiation and, correspondingly, larger values of the coefficients appearing in the exponentials. We also see in figure \ref{fig:intmscale} (b) that at high scales the expanded part approaches the fixed order $\mathrm{NNLO_{Approx.}}$, implying that, at high scales, the dominant parts of the latter stem from the large logarithms. Thus, the full NNLL matched cross section is mainly driven by $\mathrm{NNLO_{Approx.}}$ at low scales, and by the higher-order logarithms at large scales.

	We have checked that the above observations do not change for higher masses. We also note that the extremely small scale dependence observed at the ``NLO+NLL with ${\cal C}^{(1)}$'' level for $m_{\tilde{g}}=1.2$ TeV (cf. figure \ref{fig:intmscale} (a)) seems to be accidental, as it is not preserved for $m_{\tilde{g}}=2.5$ TeV.

%%%%%%%%%%%%%%%%%%%%%%%%%%%%%%%%%%%%%%%%%%%

\section{Conclusions and outlook}

%%%%%%%%%%%%%%%%%%%%%%%%%%%%%%%%%%%%%%%%%%%

\label{s:conclusion}

Using the Mellin-moment-space formalism, we have performed the NNLL resummation of soft-gluon emissions for squark and gluino hadroproduction. The resummed results are matched to the approximation of the NNLO results. The $\rm NNLO_{\rm Approx}$+NNLL predictions are then provided for LHC collisions at $\sqrt S=8$ TeV. The NNLL corrections lead to a significant increase of the size of the cross sections for all four processes of squark and gluino pair production, both with respect to the $\rm NNLO_{\rm Approx}$ and the $\rm NLO+NLL$ results. The $\rm NNLO_{\rm Approx}$+NNLL corrections are particularly important for the $\sq\sqb$ production channel where, in the mass range of up to 2.5 TeV, the enhancement over the NLO cross section can reach up to approximately 80\% at the average-mass scale of the two particles produced, compared to up to about 20\% increase over NLO due to NLL corrections. Apart from the impact of NNLL logarithmic terms and one-loop hard matching coefficients, this effect can be traced back to the inclusion of Coulomb corrections in the matching coefficients, cf.~\cite{Beenakker:2011sf, Falgari:2012hx}. Among the four processes of squark and gluino production the highest overall NNLL $K$-factor, i.e. the highest enhancement of the full matched  $\rm NNLO_{\rm Approx}$+NNLL cross section over the NLO cross section, is observed for the gluino-pair production process. In this case, however, the NNLL logarithmic terms of soft-gluon origin and the one-loop hard-matching coefficient are more important than for the $\sq\sqb$ process, see also~\cite{Beenakker:2013mva}.

Including the NNLL contributions leads to a reduction of the scale dependence for the squark and gluino production total cross sections, with the exception of gluino-pair production. This unexpected effect for the $\gl\gl$ production process can be traced back to the impact of Coulomb corrections and, more importantly, to the scale-dependent terms present in the resummed exponentials. These terms seem to spoil the compensation of the scale-dependence between the resummed expression and the evolution of the parton distributions, especially when multiplied by the matching coefficients. In this context, it is worth noting that high-mass gluino-pair production in $pp$ collisions at $\sqrt S=8$ TeV can be seen as an extreme case among the LHC processes for which the NNLL resummation has been performed so far, given the size and the importance of the logarithmic terms of soft origin that are characterized by large colour factors in the dominant gluon-gluon initial-state channel. It is also possible that the observed effect is partially caused by the incompleteness of the two-loop matching coefficient that we use here. The calculation of the complete coefficient would require knowledge of the full NNLO result, which is not yet available. Additionally, the conclusions can change upon implementing resummation of Coulomb corrections in the Mellin-moment-space approach. The impact of the resummation of Coulomb corrections is left for a future study.

While for squark-squark, squark-antisquark, and squark-gluino production our results agree relatively well with the corresponding results presented in~\cite{Beneke:2013opa}, for gluino-pair production the NNLL $K$-factors differ by around 10\%. It remains to be resolved in the future if the difference originates from the different methods of performing resummation, i.e. Mellin-moment-space approach versus effective-field-theory approach. Regarding the scale dependence of the cross section, it is difficult to perform a comparison between the two approaches as estimates of the scale uncertainty used in the effective-theory predictions cannot be directly translated into the Mellin-moment-space approach.

\section*{Acknowledgments}

We are grateful to Robert Thorne for providing us with a version of the MSTW 2008 NNLO parton distribution functions with improved accuracy in the evolution at large $x$. We thank Irene Niessen for collaboration during the early stages of this project. We also thank M. Beneke, P. Falgari, J. Piclum, Ch. Schwinn and C. Wever for correspondence. 

This work has been supported in part by the Helmholtz Alliance ``Physics at the Terascale'', the Foundation for Fundamental Research of Matter (FOM), program 104 ``Theoretical Particle Physics in the Era of the LHC", the DFG SFB/TR9 ``Computational Particle Physics'', Polish National Science Centre grant, project number DEC-2011/01/B/ST2/03643, the European Community's Marie-Curie Research Training Network under contract MRTN-CT-2006-035505 ``Tools and Precision Calculations for Physics Discoveries at Colliders'' and by the Research Executive Agency (REA) of the European Union under the Grant
Agreement number PITN-GA-2010-264564 (LHCPhenoNet).

\end{fmffile}

\appendix

\allowdisplaybreaks

\section{One-loop Coulomb corrections in N-space}

\label{app:1lcoulomb}

In this appendix we present the Mellin transforms of the LO cross sections and give the corresponding integrals for the Coulomb corrections, which receive an additional factor $1/\beta$ compared to the LO cross section. We commence by listing the integrals that are needed for calculating the Mellin transforms and subsequently give the full results for all squark and gluino production processes.

For the calculation of the Mellin transforms we use the following notation. The number of light quark flavours is denoted by $n_l$. Furthermore we define:
\[\as=\frac{g_{\rm s}^2}{4\pi}\quad\mbox{and}\quad\ashat=\frac{\hat g_{\rm s}^2}{4\pi},\]
where $g_{\rm s}$ is the QCD gauge coupling, while $\hat g_{\rm s}$ is the corresponding quark-squark-gluino coupling in the $\MSbar$ scheme. The colour-decomposed LO cross sections are given for $N_c$ colours and are labelled such that they correspond to the colour structures in Ref.~\cite{Beenakker:2013mva}. In addition we use the shorthand notations
\begin{align*}
L_1&=\log\bigg(\frac{s+2m_-^2-s\beta}{s+2m_-^2+s\beta}\bigg)\ ,&m_-^2&=m_{\tilde g}^2-m_{\tilde q}^2\ ,\\
L_2&=\log\bigg(\frac{s-2m_-^2-s\beta}{s-2m_-^2+s\beta}\bigg)\ ,&m_+^2&=m_{\tilde g}^2+m_{\tilde q}^2\ ,\\
L_3 &=\log\bigg(\frac{s+m_-^2-\kappa s\beta}{s+m_-^2+\kappa s\beta}\bigg)\ ,&\beta&=\sqrt{1-\frac{4m_{av}^2}{s}}\ ,\\
L_4 &=\log\bigg(\frac{s-m_-^2-\kappa s\beta}{s-m_-^2+\kappa s\beta}\bigg)\ ,&\kappa&= \sqrt{1-\frac{(m_\sq-m_\gl)^2}{s}}\ ,
\end{align*}
as well as
\begin{align*}
z&=\frac{4m_{av}^2}{s}=1-\beta^2,&r&=\frac{m_\gl}{m_\sq},&
A&=\frac{r-1}{r+1},&B&=\frac{r^2-1}{r^2+1},
\end{align*}
with $s$ the CM energy squared and $m_{av}$ the average mass of the produced particles.

\subsection{Integrals}

The solutions to the integrals are expressed in terms of the $\Gamma$-function:
\[\Gamma(z) = \int_0^\infty t^{z-1}e^{-t} \d t\,,\]
and the generalized hypergeometric function:
\[_pF_q(a_1,\cdots,a_p;b_1,\cdots,b_q;x)=\sum_{n=0}^\infty\frac{\Gamma(a_1\!+\!n)\cdots\Gamma(a_p\!+\!n)}{\Gamma(a_1)\cdots\Gamma(a_p)}\frac{\Gamma(b_1)\cdots\Gamma(b_q)}{\Gamma(b_1\!+\!n)\cdots\Gamma(b_q\!+\!n)}\frac{x^n}{n!}\,.\]

First we have the integrals that correspond to the linear terms in $\beta$ in the LO cross section. When including the $1/\beta$ factor from the Coulomb correction, these terms become constants. Such integrals occur in most processes and are given by:
\begin{align}
K(N)&=\int_0^1\d zz^N\sqrt{1-z}=\frac{\sqrt{\pi}\,\Gamma(N+1)}{2\,\Gamma(N+\tfrac{5}{2})}\,,\\
&\hphantom{=}\int_0^1\d zz^N=\frac{1}{N+1}\,.
\end{align}
For the quark-initiated processes we also need:
\begin{align}
K_1(N)&=\int_0^1\d z\frac{z^N\sqrt{1-z}}{z(r^2-1)^2+4r^2}=\frac{1}{(r^2+1)^2}\frac{\sqrt{\pi}\,\Gamma(N+1)}{2\,\Gamma(N+\tfrac{5}{2})}\,_2F_1(1,\tfrac{3}{2};N+\tfrac{5}{2};B^2)\,,\\
M_1(N)&=\int_0^1\d z\frac{z^N}{z(r^2-1)^2+4r^2}=\frac{1}{(r^2+1)^2(N+1)}\,_2F_1(1,1;N+2;B^2)\,.
\end{align}
For the $qg\to\sq\gl$ case we need integrals for the $\kappa\beta$ terms:
\begin{align}
K_2(N)&=\int_0^1\d zz^N\sqrt{1-z}\sqrt{1-A^2z}=\frac{\sqrt{\pi}\,\Gamma(N+1)}{2\,\Gamma(N+\tfrac{5}{2})}\,_2F_1(-\tfrac{1}{2},N+1;N+\tfrac{5}{2};A^2)\,,\\
M_2(N)&=\int_0^1\d zz^N\sqrt{1-A^2z}=\frac{1}{N+1}\,_2F_1(-\tfrac{1}{2},N+1;N+2;A^2)\,.
\end{align}
For the gluon-initiated processes, we need the additional integrals:
\begin{align}
K_3(N)&=\int_0^1\d zz^N\log\bigg(\frac{1-\sqrt{1-z}}{1+\sqrt{1-z}}\bigg)=-\frac{\sqrt{\pi}\,\Gamma(N+1)}{(N+1)\Gamma(N+\tfrac{3}{2})}\,,\\
M_3(N)&=\int_0^1\d z\frac{z^N}{\sqrt{1-z}}\log\bigg(\frac{1-\sqrt{1-z}}{1+\sqrt{1-z}}\bigg)=\frac{-2}{N+1}\,_3F_2(\tfrac{1}{2},1,1;\tfrac{3}{2},N+2;1)\,,
\end{align}
which can be calculated using the identity~\cite{integralbook}:
\begin{align*}
\log\bigg(\frac{1-\sqrt{z}}{1+\sqrt{z}}\bigg)&=-2\sqrt{z}\;\;_2F_1(\tfrac{1}{2},1;\tfrac{3}{2};z)\,.
\end{align*}
For the $q\bar q\to\sq\sqb$ process, we also need two integrals containing $L_1$. These can be obtained by rewriting the logarithm:
\begin{align*}
\log\bigg(\frac{1+\frac{1}{2}(r^2-1)z-\sqrt{1-z}}{1+\frac{1}{2}(r^2-1)z+\sqrt{1-z}}\bigg)= \log\bigg(\frac{1\!-\!\sqrt{1-z}}{1\!+\!\sqrt{1-z}}\bigg)+\log\bigg(\frac{1+B\sqrt{1-z}}{1-B\sqrt{1-z}}\bigg)\,,
\end{align*}
which leads to the solution:
\begin{align}
K_4(N,r)&=\int_0^1\d zz^N\log\bigg(\frac{1+\frac{1}{2}(r^2-1)z-\sqrt{1-z}}{1+\frac{1}{2}(r^2-1)z+\sqrt{1-z}}\bigg)\\
&\qquad= K_3(N)+B\frac{\sqrt{\pi}\,\Gamma(N+1)}{\Gamma(N+\tfrac{5}{2})}\,_2F_1(\tfrac{1}{2},1;N+\tfrac{5}{2};B^2)\,,\nonumber\\
M_4(N,r)&=\int_0^1\d z\frac{z^N}{\sqrt{1-z}}\log\bigg(\frac{1+\frac{1}{2}(r^2-1)z-\sqrt{1-z}}{1+\frac{1}{2}(r^2-1)z+\sqrt{1-z}}\bigg)\\
&\qquad= M_3(N)+\frac{2B}{N+1}\,_3F_2(\tfrac{1}{2},1,1;\tfrac{3}{2},N+2;B^2)\,.\nonumber
\end{align}
The corresponding integrals for $L_2$, which are needed for the $q\bar q\to\gl\gl$ process, can be obtained by substituting $r\to1/r$, which corresponds to $B\to-B$. 
For the $qq\to\sq\sq$ process, we also need:
\begin{align}
K_5(N,r)&=\int_0^1\d z\frac{z^N}{2+(r^2-1)z}\log\bigg(\frac{1+\frac{1}{2}(r^2-1)z-\sqrt{1-z}}{1+\frac{1}{2}(r^2-1)z+\sqrt{1-z}}\bigg)\\
&=-\frac{\Gamma(N+1)}{r^2+1}\sum_{n,k=0}^\infty\frac{B^k(1-B^{2n+1})}{n+\tfrac{1}{2}}\frac{\Gamma(n+k+\tfrac{3}{2})}{\Gamma(N+n+k+\tfrac{5}{2})}\,,\nonumber\\
&=-\frac{\Gamma(N+1)}{r^2+1}\sum_{n=0}^\infty\frac{(1-B^{2n+1})\Gamma(n+\tfrac{1}{2})}{\Gamma(N+n+\tfrac{5}{2})}\,_2F_1(1,n+\tfrac{3}{2};N+n+\tfrac{5}{2};B)\,,\nonumber\\
M_5(N,r)&=\int_0^1\d z\frac{z^N}{(2+(r^2-1)z)\sqrt{1-z}}\log\bigg(\frac{1+\frac{1}{2}(r^2-1)z-\sqrt{1-z}}{1+\frac{1}{2}(r^2-1)z+\sqrt{1-z}}\bigg)\\
&= -\frac{\Gamma(N+1)}{r^2+1}\sum_{n,k=0}^\infty\frac{B^k(1-B^{2n+1})}{n+\tfrac{1}{2}}\frac{\Gamma(n+k+1)}{\Gamma(N+n+k+2)}\nonumber\\
&=-\frac{\Gamma(N+1)}{r^2+1}\sum_{n=0}^\infty\frac{\Gamma(n+1)(1-B^{2n+1})}{(n+\tfrac{1}{2})\Gamma(n+N+2)}\,_2F_1(1,n+1;n+N+2;B)\,.\nonumber
\end{align}
Also in this case, the corresponding integrals for $L_2$, which are needed for the $q\bar q\to\gl\gl$ process, can be obtained by substituting $r\to1/r$.

For the $qg\to\sq\gl$ process, we need the Mellin transforms of $L_3$ and $L_4$:
\begin{align}
K_6^\pm(N,r)&=\int_0^1\d zz^N\log\bigg(\frac{1\pm Az-\sqrt{(1-z)(1-A^2z)}}{1\pm Az+\sqrt{(1-z)(1-A^2z)}}\bigg)\\
&=-\frac{\sqrt{\pi}\,\Gamma(N+1)}{\Gamma(N+\tfrac{3}{2})(N+1)}\,_2F_1(\tfrac{1}{2},N+1;N+\tfrac{3}{2};A^2)\nonumber\\
&\quad\pm A\frac{\sqrt{\pi}\,\Gamma(N+1)}{\Gamma(N+\tfrac{5}{2})}\,_2F_1(\tfrac{1}{2},N+2;N+\tfrac{5}{2};A^2)\,,\nonumber\\
M_6^\pm(N,r)&=\int_0^1\d z\frac{z^N}{\sqrt{1-z}}\log\bigg(\frac{1\pm Az-\sqrt{(1-z)(1-A^2z)}}{1\pm Az+\sqrt{(1-z)(1-A^2z)}}\bigg)\\ 
&=-\sum_{n=0}^\infty\frac{\Gamma(n+1)\Gamma(N+1)}{(n+\tfrac{1}{2})\Gamma(n+N+2)}\,_2F_1(n+\tfrac{1}{2},N+1;n+N+2;A^2)\nonumber\\
&\quad\pm\frac{2A}{\sqrt{1-A^2}}\frac{1}{N+1}\,_3F_2(\tfrac{1}{2},\tfrac{1}{2},1;\tfrac{3}{2},N+2;\tfrac{A^2}{A^2-1})\,.\nonumber
\end{align}

\subsection{Mellin transforms for LO Cross Sections and Coulomb Corrections}

\subsubsection[$q\bar q\to\sq\sqb$]{\boldmath$q\bar q\to\sq\sqb$}

For the $q\bar q\to\sq\sqb$ process, the Mellin transforms of the LO cross sections are:
\begin{align}
\tilde\sigma^{\rm (0)}_{q\bar q\to\sq\sqb,1}&=-\frac{\alpha_{\rm s}^2\pi(N_c^2\!-\!1)^2}{8m_\sq^2N_c^4}\bigg[2K(N)\!-4r^2K_1(N)+\!K_4(N,r)+\!\tfrac{1}{2}(r^2\!-\!1)K_4(N\!+\!1,r)\bigg]\,,\\
\tilde\sigma^{\rm (0)}_{q\bar q\to\sq\sqb,2}&=\frac{1}{N_c^2\!-\!1}\tilde\sigma^{\rm (0)}_{q\bar q\to\sq\sqb,1}+\delta_{f_1f_2}n_l\frac{\alpha_{\rm s}^2\pi(N_c^2\!-\!1)}{24m_\sq^2N_c^2}\Big(K(N)-K(N\!+\!1)\Big)\\
&\hspace{-26pt}+\delta_{f_1f_2}\!\frac{\alpha_{\rm s}^2\pi(N_c^2\!-\!1)}{8m_\sq^2N_c^3}\bigg[K(N)\!+\!\tfrac{1}{2}(r^2\!-\!1)K(N\!+\!1)\!+\!\tfrac{1}{2}r^2K_4(N\!+\!1,r)\!+\!\tfrac{1}{8}(r^2\!-\!1)^2K_4(N\!+\!2,r)\bigg]\,,\nonumber
\end{align}
while the Mellin transforms of the Coulomb corrections are given by:
\begin{align}
\tilde\sigma^{\rm Coul,(1)}_{q\bar q\to\sq\sqb,1}&=\frac{\alpha_{\rm s}^3\pi^2(N_c^2\!-\!1)^3}{32m_\sq^2N_c^5}\bigg[\frac{-2}{N\!+\!1}\!+4r^2M_1(N)-\!M_4(N,r)-\!\tfrac{1}{2}(r^2\!-\!1)M_4(N\!+\!1,r)\bigg]\,,\\
\tilde\sigma^{\rm Coul,(1)}_{q\bar q\to\sq\sqb,2}&=-\frac{1}{(N_c^2\!-\!1)^2}\tilde\sigma^{\rm Coul,(1)}_{q\bar q\to\sq\sqb,1}-\delta_{f_1f_2}n_l\frac{\alpha_{\rm s}^3\pi^2(N_c^2\!-\!1)}{96m_\sq^2N_c^3}\frac{1}{(N\!+\!1)(N\!+\!2)}\\
&\hspace{-10pt}-\delta_{f_1f_2}\frac{\alpha_{\rm s}^3\pi^2(N_c^2\!-\!1)}{32m_\sq^2N_c^4}\bigg[\frac{1}{N\!+\!1}+\frac{r^2\!-\!1}{2(N\!+\!2)}+\tfrac{1}{2}r^2M_4(N\!+\!1,r)+\tfrac{1}{8}(r^2\!-\!1)^2M_4(N\!+\!2,r)\bigg]\,.\nonumber
\end{align}

\subsubsection[$gg\to\sq\sqb$]{\boldmath$gg\to\sq\sqb$}

For the $gg\to\sq\sqb$ process, the Mellin transforms of the LO cross sections are given by:
\begin{align}
\tilde\sigma^{\rm (0)}_{gg\to\sq\sqb,1}&=\frac{\alpha_{\rm s}^2\pi n_l}{4m_\sq^2N_c(N_c^2\!-\!1)}\bigg[K(N)+K(N\!+\!1)+K_3(N\!+\!1)-\tfrac{1}{2}K_3(N\!+\!2)\bigg]\,,\\
\tilde\sigma^{\rm (0)}_{gg\to\sq\sqb,2}&=\frac{\alpha_{\rm s}^2\pi n_lN_c}{4m_\sq^2(N_c^2\!-\!1)}\bigg[\tfrac{1}{6}K(N)+\tfrac{4}{3}K(N\!+\!1)+\tfrac{1}{2}K_3(N\!+\!1)+\tfrac{1}{4}K_3(N\!+\!2)\bigg]\,,\\
\tilde\sigma^{\rm (0)}_{gg\to\sq\sqb,3}&=\tfrac{1}{2}(N_c^2\!-\!4)\,\tilde\sigma^{\rm (0)}_{gg\to\sq\sqb,1}\,,
\end{align}
and the Coulomb corrections in $N$-space are:
\begin{align}
\tilde\sigma^{\rm Coul,(1)}_{gg\to\sq\sqb,1}&=\frac{\alpha_{\rm s}^3\pi^2n_l}{16m_\sq^2N_c^2}\bigg[\frac{1}{N\!+\!1}+\frac{1}{N\!+\!2}+M_3(N\!+\!1)-\tfrac{1}{2}M_3(N\!+\!2)\bigg]\,,\\
\tilde\sigma^{\rm Coul,(1)}_{gg\to\sq\sqb,2}&=-\frac{\alpha_{\rm s}^3\pi^2n_l}{16m_\sq^2(N_c^2\!-\!1)}\bigg[\frac{1}{6(N\!+\!1)}+\frac{4}{3(N\!+\!2)}+\tfrac{1}{2}M_3(N\!+\!1)+\tfrac{1}{4}M_3(N\!+\!2)\bigg]\,,\\
\tilde\sigma^{\rm Coul,(1)}_{gg\to\sq\sqb,3}&=-\frac{1}{2}\frac{N_c^2\!-\!4}{N_c^2\!-\!1}\tilde\sigma^{\rm Coul,(1)}_{gg\to\sq\sqb,1}\,.
\end{align}

\subsubsection[$qq\to\sq\sq$]{\boldmath$qq\to\sq\sq$}

The structure of the $qq\to\sq\sq$ process has much in common with the singlet channel of the $q\bar q\to\sq\sqb$ process:
\begin{align}
\tilde\sigma^{\rm (0)}_{qq\to\sq\sq,1}&=\frac{\alpha_{\rm s}^2\pi(N_c^2\!-\!1)(N_c\!+\!1)}{16m_\sq^2N_c^3}\bigg[-\!2K(N)+4r^2K_1(N)-\!K_4(N,r)\\
&\qquad-\!\tfrac{1}{2}(r^2\!-\!1)K_4(N\!+\!1,r)+r^2\delta_{f_1f_2}K_5(N\!+\!1,r)\bigg]\,,\nonumber\\
\tilde\sigma^{\rm (0)}_{qq\to\sq\sq,2}&=\frac{\alpha_{\rm s}^2\pi(N_c^2\!-\!1)(N_c\!-\!1)}{16m_\sq^2N_c^3}\bigg[-\!2K(N)\!+4r^2K_1(N)-\!K_4(N,r)\\
&\qquad-\!\tfrac{1}{2}(r^2\!-\!1)K_4(N\!+\!1,r)-r^2\delta_{f_1f_2}K_5(N\!+\!1,r)\bigg]\,,\nonumber
\end{align}
and the Mellin transforms of the Coulomb corrections are:
\begin{align}
\tilde\sigma^{\rm Coul,(1)}_{qq\to\sq\sq,1}&=\frac{\alpha_{\rm s}^3\pi^2(N_c^2\!-\!1)(N_c\!+\!1)^2}{64m_\sq^2N_c^4}\bigg[\frac{-2}{N\!+\!1}\!+4r^2M_1(N)-\!M_4(N,r)\\
&\qquad-\!\tfrac{1}{2}(r^2\!-\!1)M_4(N\!+\!1,r)+r^2\delta_{f_1f_2}M_5(N\!+\!1,r)\bigg]\,,\nonumber\\
\tilde\sigma^{\rm Coul,(1)}_{qq\to\sq\sq,2}&=\frac{\alpha_{\rm s}^3\pi^2(N_c^2\!-\!1)(N_c\!-\!1)^2}{64m_\sq^2N_c^4}\bigg[\frac{2}{N\!+\!1}\!-4r^2M_1(N)+\!M_4(N,r)\\
&\qquad+\!\tfrac{1}{2}(r^2\!-\!1)M_4(N\!+\!1,r)+r^2\delta_{f_1f_2}M_5(N\!+\!1,r)\bigg]\,.\nonumber
\end{align}

\subsubsection[$q\bar q\to\gl\gl$]{\boldmath$q\bar q\to\gl\gl$}

The Mellin transforms of the LO cross sections of the $q\bar q\to\gl\gl$ process are given by:
\begin{align}
\tilde\sigma^{\rm (0)}_{q\bar q\to\gl\gl,1}&=\frac{\alpha_{\rm s}^2\pi(N_c^2\!-\!1)}{8m_\gl^2N_c^3}\bigg[2K(N)-4r^2K_1(N)-\frac{r^2\!-\!1}{2r^2}K_4(N\!+\!1,\tfrac{1}{r})\\
&\qquad+K_5(N\!+\!1,\tfrac{1}{r})\bigg]\,,\nonumber\\
\tilde\sigma^{\rm (0)}_{q\bar q\to\gl\gl,2}&=\frac{\alpha_{\rm s}^2\pi(N_c^2\!-\!1)}{16m_\gl^2N_c}\bigg[\frac{3\!-\!r^2}{3r^2}K(N\!+\!1)-4r^2K_1(N)+\tfrac{4}{3}K(N)\\
&\qquad+\frac{r^2\!+\!1}{2r^2}K_4(N\!+\!1,\tfrac{1}{r})+\frac{(r^2\!-\!1)^2}{4r^4}K_4(N\!+\!2,\tfrac{1}{r})-K_5(N\!+\!1,\tfrac{1}{r})\bigg]\,,\nonumber\\
\tilde\sigma^{\rm (0)}_{q\bar q\to\gl\gl,3}&=\tfrac{1}{2}(N_c^2-4)\,\tilde\sigma^{\rm (0)}_{q\bar q\to\gl\gl,1}\,,
\end{align}
and the Mellin-transformed Coulomb corrections are:
\begin{align}
\tilde\sigma^{\rm Coul,(1)}_{q\bar q\to\gl\gl,1}&=\frac{\alpha_{\rm s}^3\pi^2(N_c^2\!-\!1)}{16m_\gl^2N_c^2}\bigg[\frac{2}{N\!+\!1}-4r^2M_1(N)-\frac{r^2\!-\!1}{2r^2}M_4(N\!+\!1,\tfrac{1}{r})\\
&\qquad+M_5(N\!+\!1,\tfrac{1}{r})\bigg]\,,\nonumber\\
\tilde\sigma^{\rm Coul,(1)}_{q\bar q\to\gl\gl,2}&=\frac{\alpha_{\rm s}^3\pi^2(N_c^2\!-\!1)}{64m_\gl^2}\bigg[\frac{3\!-\!r^2}{3r^2}\frac{1}{N\!+\!2}-4r^2M_1(N)+\frac{4}{3(N\!+\!1)}\\
&\qquad+\frac{r^2\!+\!1}{2r^2}M_4(N\!+\!1,\tfrac{1}{r})+\frac{(r^2\!-\!1)^2}{4r^4}M_4(N\!+\!2,\tfrac{1}{r})-M_5(N\!+\!1,\tfrac{1}{r})\bigg]\,,\nonumber\\
\tilde\sigma^{\rm Coul,(1)}_{q\bar q\to\gl\gl,3}&=\tfrac{1}{4}(N_c^2-4)\,\tilde\sigma^{\rm Coul,(1)}_{q\bar q\to\gl\gl,1}\,.
\end{align}

\subsubsection[$gg\to\gl\gl$]{\boldmath$gg\to\gl\gl$}

For the $gg\to\gl\gl$ process, the Mellin transforms of the LO cross sections are given by:
\begin{align}
\hspace{-1.2ex}\tilde\sigma^{\rm (0)}_{gg\to\gl\gl,1}&=\frac{-\alpha_{\rm s}^2\pi N_c^2}{2m_\gl^2(N_c^2\!-\!1)^2}\bigg[K(N)+K(N\!+\!1)+K_3(N)+K_3(N\!+\!1)-\tfrac{1}{2}K_3(N\!+\!2)\bigg]\,,\\
\hspace{-1.2ex}\tilde\sigma^{\rm (0)}_{gg\to\gl\gl,2}&=\frac{-\alpha_{\rm s}^2\pi N_c^2}{8m_\gl^2(N_c^2\!-\!1)}\bigg[\tfrac{7}{3}K(N)+\tfrac{8}{3}K(N\!+\!1)+K_3(N)+K_3(N\!+\!1)+\tfrac{1}{2}K_3(N\!+\!2)\bigg]\,,\\
\hspace{-1.2ex}\tilde\sigma^{\rm (0)}_{gg\to\gl\gl,3}&=\tfrac{1}{4}(N_c^2\!-\!1)\,\tilde\sigma^{\rm (0)}_{gg\to\gl\gl,1}\,,\\
\hspace{-1.2ex}\tilde\sigma^{\rm (0)}_{gg\to\gl\gl,4}&=\tilde\sigma^{\rm (0)}_{gg\to\gl\gl,5}=0\,,\\
\hspace{-1.2ex}\tilde\sigma^{\rm (0)}_{gg\to\gl\gl,6}&=\tfrac{1}{4}(N_c\!+\!3)(N_c\!-\!1)\,\tilde\sigma^{\rm (0)}_{gg\to\gl\gl,1}\,,\\
\hspace{-1.2ex}\tilde\sigma^{\rm (0)}_{gg\to\gl\gl,7}&=\tfrac{1}{4}(N_c\!-\!3)(N_c\!+\!1)\,\tilde\sigma^{\rm (0)}_{gg\to\gl\gl,1}\,,
\end{align}
and the Mellin transforms of the Coulomb corrections are:
\begin{align}
\tilde\sigma^{\rm Coul,(1)}_{gg\to\gl\gl,1}&=\frac{-\alpha_{\rm s}^3\pi^2N_c^3}{4m_\gl^2(N_c^2\!-\!1)^2}\bigg[\frac{1}{N\!+\!1}+\frac{1}{N\!+\!2}+M_3(N)+M_3(N\!+\!1)-\tfrac{1}{2}M_3(N\!+\!2)\bigg]\,,\\
\tilde\sigma^{\rm Coul,(1)}_{gg\to\gl\gl,2}&=\frac{-\alpha_{\rm s}^3\pi^2N_c^3}{32m_\gl^2(N_c^2\!-\!1)}\bigg[\frac{7}{3(N\!+\!1)}+\frac{8}{3(N\!+\!2)}+M_3(N)+M_3(N\!+\!1)\\
&\qquad+\tfrac{1}{2}M_3(N\!+\!2)\bigg]\,,\nonumber\\
\tilde\sigma^{\rm Coul,(1)}_{gg\to\gl\gl,3}&=\tfrac{1}{8}(N_c^2\!-\!1)\tilde\sigma^{\rm Coul,(1)}_{gg\to\gl\gl,1}\,,\\
\tilde\sigma^{\rm Coul,(1)}_{gg\to\gl\gl,4}&=\tilde\sigma^{\rm Coul,(1)}_{gg\to\gl\gl,5}=0\,,\\
\tilde\sigma^{\rm Coul,(1)}_{gg\to\gl\gl,6}&=-\frac{(N_c\!+\!3)(N_c\!-\!1)}{4N_c}\tilde\sigma^{\rm Coul,(1)}_{gg\to\gl\gl,1}\,,\\
\tilde\sigma^{\rm Coul,(1)}_{gg\to\gl\gl,7}&=\frac{(N_c\!-\!3)(N_c\!+\!1)}{4N_c}\tilde\sigma^{\rm Coul,(1)}_{gg\to\gl\gl,1}\,.
\end{align}

\subsubsection[$qg\to\sq\gl$]{\boldmath$qg\to\sq\gl$}

The Mellin transforms of the LO cross sections of the $qg\to\sq\gl$ process are given by:
\begin{align}
\tilde\sigma^{\rm (0)}_{qg\to\sq\gl,1}&=\frac{\alpha_{\rm s}^2\pi}{4m_{av}^2(N_c^2\!-\!1)}\bigg[\bigg(\frac{1}{2}+\frac{1}{4N_c^2}-\frac{3N_c^2}{4}\bigg)K_2(N)-\frac{N_c^2}{2}K_6^+(N)\\
&\quad+\bigg(\frac{3}{2}-\frac{7}{4N_c^2}-\frac{7N_c^2}{4}\bigg)AK_2(N\!+\!1)+\frac{A(1\!+\!A)^2}{2}K_6^+(N\!+\!2)-AN_c^2K_6^+(N\!+\!1)\nonumber\\
&\quad-A^2N_c^2K_6^+(N\!+\!2)+\frac{A(1\!-\!A)^2}{2}K_6^-(N\!+\!2)-\frac{A}{N_c^2}K_6^-(N\!+\!1)+\frac{A^2}{N_c^2}K_6^-(N\!+\!2)\bigg]\,,\nonumber\\
\tilde\sigma^{\rm (0)}_{qg\to\sq\gl,2}&=\frac{\alpha_{\rm s}^2\pi(N_c\!-\!2)}{4m_{av}^2(N_c\!-\!1)}\bigg[\frac{A(1\!+\!A^2)}{4}K_6^-(N\!+\!2)-AK_2(N\!+\!1)\\
&\qquad-\frac{A}{2}K_6^-(N\!+\!1)-\frac{1}{4}K_6^+(N)-\frac{A}{2}K_6^+(N\!+\!1)+\frac{A}{4}(1\!+\!A^2)K_6^+(N\!+\!2)\bigg]\,,\nonumber\\
\tilde\sigma^{\rm (0)}_{qg\to\sq\gl,3}&=\frac{N_c\!+\!2}{N_c\!+\!1}\frac{N_c\!-\!1}{N_c\!-\!2}\tilde\sigma^{\rm (0)}_{qg\to\sq\gl,2}\,,
\end{align}
and the Coulomb corrections in $N$-space are:
\begin{align}
\tilde\sigma^{\rm Coul,(1)}_{qg\to\sq\gl,1}&=\frac{\alpha_{\rm s}^3\pi^2N_c}{16m_{av}^2(N_c^2\!-\!1)}\sqrt{\frac{m_\gl m_\sq}{m_{av}^2}}\bigg[\bigg(\frac{1}{2}+\frac{1}{4N_c^2}-\frac{3N_c^2}{4}\bigg)M_2(N)-\frac{N_c^2}{2}M_6^+(N)\\
&\quad+\bigg(\frac{3}{2}-\frac{7}{4N_c^2}-\frac{7N_c^2}{4}\bigg)AM_2(N\!+\!1)+\frac{A(1\!+\!A)^2}{2}M_6^+(N\!+\!2)-AN_c^2M_6^+(N\!+\!1)\nonumber\\
&\quad-A^2N_c^2M_6^+(N\!+\!2)+\frac{A(1\!-\!A)^2}{2}M_6^-(N\!+\!2)-\frac{A}{N_c^2}M_6^-(N\!+\!1)+\frac{A^2}{N_c^2}M_6^-(N\!+\!2)\bigg]\,,\nonumber\\
\tilde\sigma^{\rm Coul,(1)}_{qg\to\sq\gl,2}&=\frac{\alpha_{\rm s}^3\pi^2(N_c\!-\!2)}{16m_{av}^2(N_c\!-\!1)}\sqrt{\frac{m_\gl m_\sq}{m_{av}^2}}\bigg[\frac{A(1\!+\!A^2)}{4}M_6^-(N\!+\!2)-AM_2(N\!+\!1)\\
&\qquad-\frac{A}{2}M_6^-(N\!+\!1)-\frac{1}{4}M_6^+(N)-\frac{A}{2}M_6^+(N\!+\!1)+\frac{A}{4}(1\!+\!A^2)M_6^+(N\!+\!2)\bigg]\,,\nonumber\\
\tilde\sigma^{\rm Coul,(1)}_{qg\to\sq\gl,3}&=-\frac{N_c\!+\!2}{N_c\!+\!1}\frac{N_c\!-\!1}{N_c\!-\!2}\tilde\sigma^{\rm Coul,(1)}_{qg\to\sq\gl,2}\,.
\end{align}

\section{Two-loop Coulomb corrections in N-space}
\label{app:2lcoulomb}

In this appendix we give the expression for the Mellin transform of the near-threshold approximation of the two-loop Coulomb correction. The second-order Coulomb corrections for equal-mass final-state particles is taken from~\cite{Beneke:2009ye}. For the case of unequal masses for the final-state particles, we have derived the expression by expanding the imaginary part of the Green's function in Ref.~\cite{Kauth:2011bz} in terms of $\beta$ with $v\approx\beta\sqrt{m_{av}/\left(2m_{red}\right)}$. Taking the Mellin transform of the contribution of second order in $\alpha_{\rm s}$ results in:
\begin{eqnarray}
\frac{\alpha_{\rm s}^{2}\mathcal{C}^{coul,\left(2\right)}}{\pi^{2}} & = & {\displaystyle \frac{\alpha_{\rm s}^{2}}{\left(4\pi\right)^{2}}\left\{ \frac{8}{3}\left(\kappa_{ij\rightarrow kl,I}\right)^{2}\pi^{4}N\frac{2m_{red}}{m_{av}}+\kappa_{ij\rightarrow kl,I}\pi^{2}\sqrt{\frac{N}{\pi}}\sqrt{\frac{2m_{red}}{m_{av}}}\left[\vphantom{\frac{A}{A}}-16\pi b_{0}\left(\log\left(N\right) \right.\right.\right.}\nonumber \\
 &  & {\displaystyle \left. \left. + \gamma_{E}\right) -4a_{1}+32\pi b_{0}\log\left(2\right)-16\pi b_{0}\log\left(\frac{\mu^{2}}{m_{av}^{2}}\right)+16\pi b_{0}\log\left(\frac{2m_{red}}{m_{av}}\right)\right]}\nonumber \\
 &  & {\displaystyle +16\pi^{2}\kappa_{ij\rightarrow kl,I}\left(C_{A}-2\kappa_{ij\rightarrow kl,I}\left(1+v_{spin}\right)\right)\left(1-\log\left(2\right)-\frac{1}{2}\log\left(N\right)-\frac{1}{2}\gamma_{E}\right)}\nonumber \\
 &  & {\displaystyle \left.+\frac{4}{3}\left(\kappa_{ij\rightarrow kl,I}\right)^{2}\pi^{4}\frac{2m_{red}}{m_{av}}\right\} }
\label{eq:Coul2l}
\end{eqnarray} 
with $b_0=(11C_A-2n_l)/12\pi$, $a_1=(31C_A-10n_l)/9$, $C_A=N_c$ and $n_l$ being the number of light quark flavours. Furthermore, $\gamma_E$ stands for Euler's gamma constant. For two final-state particles $k$ and $l$ with masses $m_k$ and $m_l$, the reduced and average masses are defined as $m_{red}=m_km_l/(m_k+m_l)$ and $m_{av}=(m_k+m_l)/2$, correspondingly. The $\kappa$-coefficients are given by $\kappa_{ij\rightarrow kl,I}=(C_2(I)-C_k-C_l)/2$ with $C_2(I)$ the quadratic Casimir invariant of the representation $I$ and $C_k$ the colour factor of particle $k$. Note that in our approach the ${\cal O}(\alpha_{\rm s}^2)$ non-Coulomb contribution including the relativistic kinetic-energy correction~\cite{Beneke:2009ye} is also incorporated into ${C}^{coul,\left(2\right)}$. The values of $v_{spin}$ are taken from~\cite{Beneke:2013opa}.

\bibliographystyle{JHEP}
\providecommand{\href}[2]{#2}\begingroup\raggedright\endgroup

\end{document}